\documentclass[pra,twocolumn,showkeywords,longbibliography,superscriptaddress,10pt]{revtex4-1}

\usepackage{amsmath}
\usepackage{braket} 
\usepackage{graphicx}
\usepackage{dcolumn}
\usepackage{color}
\usepackage{times}
\usepackage{bm}
\usepackage{amssymb}
\usepackage{amsmath}
\usepackage{epsfig}
\usepackage{epstopdf}
\usepackage{dsfont}
\usepackage{subfigure}
\usepackage{tikz}
\usepackage[colorlinks, citecolor=blue, linkcolor=blue,urlcolor=blue]{hyperref}
\usepackage[mathscr]{euscript}
\usepackage{amsthm}
\makeatletter
\renewcommand{\maketag@@@}[1]{\hbox{\m@th\normalsize\normalfont#1}}
\makeatother
\usepackage{threeparttable}
\usepackage{booktabs}
\usepackage{tabularx}
\usepackage{array}        
\usepackage{booktabs}     
\usepackage{multirow}     
\usepackage{xcolor}       
\usepackage{ragged2e}     
\usepackage{tabularx}
\makeatletter
\@ifpackagelater{array}{2024/05/31}{\let\switch@array\relax}{}
\makeatother
\begin{document}

	\title{High-precision measurement of level leakage via intelligent coherent control}
\author{Ting Wang}\email{These authors contributed equally.} 
\affiliation{School of Physics, Anhui University, Hefei 230601,  China}
\author{Hang Xu}\email{These authors contributed equally.}
\affiliation{State Key Laboratory of Photonics and Communications, Institute for Quantum Sensing and Information Processing, Shanghai Jiao Tong University, Shanghai 200240, China}
\author{Shi-Qiang Qiao}
\affiliation{School of Physics, Anhui University, Hefei 230601,  China}
\author{Xue-Ke Song} \email{songxk@ahu.edu.cn}
\affiliation{School of Physics, Anhui University, Hefei 230601,  China}
\author{Haipeng Cao}
\affiliation{School of Computer Science and Technology, University of Science and Technology of China, Hefei 230026, China}
\author{Zhao-Feng Su} \email{youngpath2012@gmail.com}
\affiliation{Research Institute of Quantum Technology, The Hong Kong Polytechnic University, Hong Kong}
	\author{Liu Ye}
\affiliation{School of Physics, Anhui University, Hefei 230601,  China}
	\author{Dong Wang}
\affiliation{School of Physics, Anhui University, Hefei 230601,  China}

\date{\today }

\begin{abstract}

Level leakage constitutes a dominant error source in quantum state control and necessitates accurate characterization for large-scale quantum computing. Quantum sensing provides a viable route for high-precision estimation of level leakage, but its sensitivity can be limited by non-unitary leakage dynamics and interference between the free Hamiltonian and signal generator. This work proposes a reinforcement learning-based quantum sensing algorithm for Hermitian three-level and effective dissipative two-level systems. The algorithm adopts an experimentally feasible pulse-parameter
	modulation strategy, enabling high-precision and reliable measurement
	under the specified control and measurement protocol. In both models,
	DQN-discovered pulse sequences enhance the measurement-model-specific
	sensing precision, while joint phase--detuning modulation further
	improves the performance and yields a CFI close to the corresponding QFI. 
\end{abstract}

\maketitle

\section{Introduction}
\label{sec1}
	
Level leakage represents a primary source of error in quantum devices \cite{PhysRevA.97.032306,Miao2023,PhysRevLett.134.120601,PhysRevA.97.032306,PhysRevA.111.032619}. Unlike Pauli errors confined to the computational subspace, leakage
transfers population outside that subspace. When only the computational
subspace is retained, this process can be described by effective
trace-decreasing dynamics. \cite{PhysRevApplied.14.034003,4kkq-x5km,Wu2025,PhysRevLett.123.100501,lycw-g48q}.
Techniques developed in quantum coherent control, such as composite pulses \cite{PhysRevA.80.043408,PhysRevA.80.013417,PhysRevA.97.043408,PhysRevLett.106.233001,PhysRevApplied.21.044012} and shortcuts to adiabaticity \cite{PhysRevA.68.043401,Xu2023,Guang-Hui,PhysRevA.102.062603}, can suppress leakage errors for specific quantum control tasks. However, these techniques are inherently task-dependent: they are designed for particular control objectives or dynamical trajectories and generally cannot be transferred or reused for other quantum computing tasks.
Quantum error correction and error mitigation schemes can, in principle, eliminate leakage at the logical qubit level, thereby providing a more general paradigm \cite{PhysRevLett.119.180501,PhysRevLett.131.210602,PhysRevLett.119.180509,PhysRevA.103.042605,Takagi2022,Quek2024,PhysRevA.109.022410}. However, such approaches rely on precise characterization of the leakage strength; otherwise, imperfect correction may introduce additional errors instead of suppressing them. Therefore, in the era of fault-tolerant quantum computing, accurately estimating device-level leakage noise is an essential prerequisite for reliable real-time quantum computation.

Quantum metrology offers a powerful framework for high-precision parameter estimation
\cite{DiFresco2025,Yuan2017,Pang2017,RevModPhys.90.035006,RevModPhys.90.035005}.
Under ideal unitary conditions, Ramsey interferometry
	provides a standard reference in which the quantum Fisher information
	can scale quadratically with the total interrogation time
	\cite{PhysRevLett.96.010401,PhysRevLett.107.083601,Gietka_2021}.
	However, for level-leakage estimation, the attainable information
	scaling depends on the physical model, control protocol, and measurement
	scheme. Coupling to leakage levels, nonunitary dynamics in effective
	dissipative descriptions, and noncommutativity between the system
	Hamiltonian and the leakage signal generator can modify coherent
	information accumulation and make the ideal quadratic scaling
	nontrivial to attain \cite{9mjc-nhxx}.
First, depending on the physical description, leakage can involve
coherent population transfer to explicitly retained leakage levels or
effective dissipative loss from the retained Hilbert space. In the
latter case, the reduced evolution is nonunitary and trace-decreasing \cite{PhysRevLett.131.160801,PhysRevA.111.022220,doi:10.1126/sciadv.adk7616,Naikoo_2026,PhysRevA.82.052341,gmcm-mlhk}. Second, the Hamiltonian of the system generally does not commute with the signal generator associated with leakage, preventing the system from remaining on the optimal trajectory in Hilbert space during free evolution \cite{PhysRevLett.81.2594,PhysRevLett.102.080405}.
For the former limitation, the total leakage probability can be recovered a posteriori from the known number of measurement trials. This situation commonly arises in optical platforms, where the total number of emitted photons is known a priori. For the latter limitation, the non-commutativity of the signal generator can be addressed through quantum control.
Gate-based quantum control strategies discretize the sensing process into time slices and insert control operations between them. However, such methods are experimentally challenging, as they require interrupting the free evolution of the system and introduce additional time overhead. An alternative approach is pulse-level parameter modulation, which directly modifies the parameters of the system Hamiltonian and does not require halting free evolution \cite{PhysRevResearch.3.L032066,PhysRevLett.132.193801,kh36-7z76}. This approach is well suited to current optical and superconducting platforms. Nevertheless, regardless of the specific control strategy employed, identifying the optimal control protocol remains a highly nontrivial task.

Optimal control techniques play central roles in different quantum information processing tasks, such as quantum sensing and quantum computation. Traditional methods for searching optimal control sequences, such as gradient ascent pulse engineering (GRAPE) \cite{Khaneja2005OptimalCO,PRXQuantum.2.010101,PhysRevA.79.053417,PhysRevA.84.022305,PhysRevA.90.032329} and chopped random basis (CRAB) \cite{PhysRevA.84.022326,PhysRevLett.106.190501,DEFOUQUIERES2011412,PhysRevA.98.022119,PhysRevResearch.4.013133,PhysRevLett.131.010801}, perform well in low-dimensional settings. However, as the control landscape grows, these methods can become
computationally expensive and may converge slowly, particularly when the
control sequence is long or the control parameters are strongly coupled. Deep reinforcement learning (DRL), leveraging the expressive power of deep neural networks, offers a promising alternative \cite{PhysRevLett.134.120803,PhysRevLett.107.233601,Knott_2016,PhysRevLett.123.230502,Xiao_2023,PhysRevX.11.031070,PhysRevLett.113.043001,PhysRevA.108.032616,PhysRevX.8.031084}. In recent years, reinforcement learning has demonstrated remarkable success in quantum computing, including quantum error-correcting code design, many-body ground-state preparation, and digital quantum simulation. Therefore, reinforcement learning provides a flexible numerical
approach for searching for high-quality pulse-control protocols that
enhance the measurement-model-specific CFI under the specified physical
models and control constraints.

In this work, we develop a reinforcement-learning-based quantum sensing framework for precision estimation of level leakage in a Hermitian three-level model and an effective non-Hermitian two-level model. We implement experimentally motivated pulse-level parameter modulation within the reinforcement-learning framework and use the classical Fisher information (CFI) to quantify the sensitivity achieved under the specified measurement protocols. The results show that pulse sequences optimized by a deep Q-network (DQN) enhance the measurement-model-specific CFI under the specified control and measurement conditions.  The joint phase--detuning control strategy further improves the sensing
	precision in both model systems, with the resulting CFI closely
	approaching their corresponding QFI. These results suggest that the proposed framework may provide a useful approach for leakage characterization in near-term
	quantum-sensing experiments and motivate further studies of leakage-aware control for fault-tolerant quantum information processing.
    
 The paper is organized as follows. In Sec. \ref{sec2}, we introduce the theoretical framework of quantum sensing and the Fisher information employed to quantify parameter-estimation sensitivity. In Sec.~\ref{sec3}, we develop a reinforcement learning framework for pulse-level quantum control and describe the implementation of a DQN-based optimization algorithm. In Sec. \ref{sec4}, we demonstrate the sensing performance of the proposed approach in a three-level leakage system and extend it to an effective non-Hermitian two-level model capturing attenuation dynamics. The robustness and practical feasibility of the learned control protocols, including training stability, experimental imperfections, and finite-shot measurement effects, are also analyzed in this section. Finally, Sec. \ref{sec5} summarizes our results and discusses potential directions for future research.

\section{Fundamental theory of quantum Sensing}
\label{sec2}

Quantum sensing aims to estimate an unknown physical parameter by exploiting the parameter-dependent evolution of a quantum system. In a general setting, the unknown parameter $\theta$ is encoded into a quantum state $\hat{\rho}_\theta$ through either unitary dynamics, dissipative dynamics, or their combined evolution. The achievable estimation precision is fundamentally limited by the amount of parameter information contained in $\hat{\rho}_\theta$ and by the measurement scheme used to extract this information.

For $M$ independent measurements and a locally unbiased estimator
$\hat{\theta}$, the quantum Cram\'er--Rao bound (QCRB) takes the form
\cite{PhysRevLett.72.3439,Paris2008QuantumEF,Liu_2020}
\begin{equation}
	\operatorname{Var}(\hat{\theta})
	\geq
	\frac{1}{M\mathcal{F}_{Q}[\hat{\rho}_{\theta}]}.
\end{equation}
Here, $M$ denotes the number of independent measurements, and
$\mathcal{F}_{Q}[\hat{\rho}_{\theta}]$ is the quantum Fisher information (QFI).
The QFI quantifies the intrinsic sensitivity of a normalized quantum
state to infinitesimal changes in the parameter. Under the usual
regularity conditions, it provides a measurement-independent upper bound
on the CFI attainable with a specified complete measurement.

For a normalized pure state $|\psi_\theta\rangle$, the QFI is given by
\begin{equation}
    \mathcal{F}_Q[|\psi_\theta\rangle]
    =
    4\left(
    \langle\partial_\theta\psi_\theta|\partial_\theta\psi_\theta\rangle
    -
    |\langle\psi_\theta|\partial_\theta\psi_\theta\rangle|^2
    \right).
    \label{eq:qfi_pure}
\end{equation}
In numerical simulations, instead of evaluating the derivative in Eq. \eqref{eq:qfi_pure} directly, we estimate the QFI from the distinguishability between two nearby quantum states, $|\psi(\theta)\rangle$ and $|\psi(\theta+\delta\theta)\rangle$. For pure states, their fidelity is
\begin{equation}
    F_U(\theta,\theta+\delta\theta)
    =
    |\langle\psi(\theta)|\psi(\theta+\delta\theta)\rangle|^2.
\end{equation}
Equivalently, using the absolute overlap
\begin{equation}
    F(\theta,\theta+\delta\theta)
    =
    |\langle\psi(\theta)|\psi(\theta+\delta\theta)\rangle|,
\end{equation}
the Bures distance is
\begin{equation}
    d_B^2
    =
    2\left(1-F\right).
\end{equation}
In the limit $\delta\theta\rightarrow 0$, the QFI is related to the Bures distance by
\begin{equation}
    d_B^2
    \simeq
    \frac{1}{4}\mathcal{F}_Q(\delta\theta)^2.
\end{equation}
Therefore, the QFI can be evaluated as
\begin{equation}
    \mathcal{F}_Q[|\psi_\theta\rangle]
    =
    \lim_{\delta\theta\rightarrow 0}
    \frac{8\left(1-
    |\langle\psi(\theta)|\psi(\theta+\delta\theta)\rangle|
    \right)}
    {(\delta\theta)^2}.
\end{equation}
In the discrete-time simulations conducted in this work, we adopt the finite-difference approximation to obtain
\begin{equation}
    \mathcal{F}_Q[|\psi_\theta\rangle]
    \approx
    \frac{8\left(1-F\right)}{(\delta\theta)^2}.
\end{equation}

This expression shows that a faster decrease of the fidelity between two nearby parameter-dependent states corresponds to a larger QFI and hence to a higher intrinsic parameter sensitivity.

The QFI characterizes the information encoded in a normalized quantum
 	state, whereas the experimentally extracted information depends on the
 	chosen measurement. For a complete positive operator-valued measure
 	(POVM) $\{\hat{\Pi}_x\}$ satisfying
 	\[
 	\sum_x\hat{\Pi}_x=\mathbb{I},
 	\]
 	the probability of obtaining outcome $x$ is
 	\[
 	p(x|\theta)=\mathrm{Tr}[\hat{\Pi}_x\hat{\rho}_\theta].
 	\]
 	The corresponding CFI is
 	\begin{equation}
 		\mathcal{F}_C
 		=
 		\sum_x
 		\frac{1}{p(x|\theta)}
 		\left(
 		\frac{\partial p(x|\theta)}{\partial\theta}
 		\right)^2.
 	\end{equation}
 	For any specified complete POVM applied to a normalized physical state,
 	the corresponding CFI satisfies
 	\begin{equation}
 		\mathcal{F}_C\leq\mathcal{F}_Q.
 	\end{equation}
 	Thus, the QFI provides a measurement-independent upper bound on the CFI,
 	whereas the CFI quantifies the information extracted under the specified
 	measurement scheme. When the measurement optimization is admissible and
 	the usual regularity conditions hold, the QFI bound can be attained.  In the discrete-time control protocol, the total sensing duration is
 	divided into $N$ equal time slices, each with duration $\Delta t$.
 	Therefore, the total sensing time is given by $T=N\Delta t.$

\section{Deep Reinforcement Learning-Based Quantum Control Framework}
\label{sec3}

To address the challenges in energy-level leakage sensing, we construct a quantum control framework based on deep reinforcement learning. This framework systematically integrates the design of the state, action, and reward, together with an efficient training algorithm incorporating experience replay, target networks, and exploration strategies.

\subsection{Basic Theoretical Framework}

We consider a general parameter-dependent quantum system governed by a controlled Hamiltonian
\begin{equation}
    H(\theta;\mathbf{u}_t),
\end{equation}
where $\mathbf{u}_t$ denotes the set of externally tunable pulse parameters at time step $t$. These pulse parameters may include, for example, the pulse phase, amplitude, detuning, duration, or other experimentally adjustable control degrees of freedom. Under a sequence of pulse controls selected by the DRL agent, the system evolves according to the Schr\"odinger equation. This abstract formulation allows the learning framework to be introduced independently of a specific microscopic Hamiltonian.

The state of a quantum system is fully described by its wave function. To interface the quantum dynamics with a classical neural network, we map the complex-valued Hilbert space to a real vector space. Specifically, we simultaneously evolve two nearby quantum states: the nominal state $|\psi(\theta)\rangle$ and the perturbed state $|\psi(\theta+d\theta)\rangle$. The comparison between these two trajectories enables real-time evaluation of the parameter sensitivity of the quantum evolution, thereby providing the information required for the calculation of the CFI.

For a finite-dimensional quantum system with Hilbert-space dimension $d$, each quantum state contains $d$ complex amplitudes. Consequently, the reinforcement-learning state vector $s_t$ is constructed by concatenating the real and imaginary parts of the nominal and perturbed states,
\begin{equation}
    s_t =
    \big[
    \mathrm{Re}(\psi_\theta),
    \mathrm{Im}(\psi_\theta),
    \mathrm{Re}(\psi_{\theta+d\theta}),
    \mathrm{Im}(\psi_{\theta+d\theta})
    \big],
\end{equation}
  
which gives a $4d$-dimensional real representation. This encoding
	preserves the quantum-state information required for evaluating the
	parameter sensitivity and constructing the reinforcement-learning
	reward. The unknown parameter $\eta$ is not provided as an input to the
	DRL agent.

The action space is defined by discretizing the experimentally tunable pulse parameters contained in $\mathbf{u}_t$. In the one-dimensional setting, the agent controls a single pulse parameter, such as the pulse phase. In the two-dimensional setting, two pulse parameters are discretized and jointly optimized, for example the pulse phase and detuning. More generally, the action at each time step can be written as
\begin{equation}
    a_t \in \mathcal{A}
    =
    \left\{
    \mathbf{u}^{(1)}, \mathbf{u}^{(2)}, \ldots, \mathbf{u}^{(N_a)}
    \right\},
\end{equation}
where $\mathcal{A}$ denotes the finite set of available pulse-control configurations. This design balances controllability and computational complexity while maintaining experimental feasibility.

The reward function guides the agent to learn a control policy that improves parameter-estimation precision. Since the two physical models exhibit different information-growth behaviors, we adopt model-specific reward functions. For Case A, the instantaneous reward is defined as
 	\begin{equation}
 		r_t^{(A)}=\log\left(\frac{F_C(t)}{t^2}\right),
 	\end{equation}
 	whereas for Case B, the reward is defined as the increment in the logarithm of the CFI between two consecutive time steps:
 	\begin{equation}
 		\begin{aligned}
 			r_t^{(B)}
 			&=\Delta\log\left[1+F_C(t)\right].\\
 		\end{aligned}
 	\end{equation}
 	Here, $F_C(t)$ denotes the classical Fisher information evaluated after the first $t$ control pulses, with $t=1,\ldots,N$. The difference operator $\Delta$ denotes the difference between consecutive time steps. The term $1+F_C(t)$ ensures that the logarithm remains well-defined when the CFI vanishes.

\begin{table*}
    \renewcommand{\thetable}{1}
    \caption{DQN-based pulse-control algorithm for quantum sensing}
    
    \small
    \centering
    
    \noindent\fbox{%
        \begin{minipage}{\dimexpr\linewidth-2\fboxsep-2\fboxrule\relax}
            \renewcommand{\arraystretch}{1.2}
            \vspace{5pt}
            
            \textbf{Input}
            
            \begin{itemize}
                \item[] Maximum evolution time $T_{\max}$, number of training episodes
                $N_{\mathrm{eps}}$, learning rate $l_r$, batch size $n_{\mathrm{batch}}$,
                discount factor $\gamma$, replay-buffer size $N_{\mathrm{pool}}$, initial
                exploration rate $\varepsilon_1$, and final exploration rate
                $\varepsilon_N$.
            \end{itemize}
            
            \vspace{8pt}
            \textbf{Initialization}
            
            \begin{itemize}
                \item[1.] \textbf{Initialize the online Q-network $Q(s,a;w)$:} 
                A deep neural network, e.g., a multi-layer perceptron, used for action-value estimation.
                
                \item[2.] \textbf{Initialize the target network:} 
                Set $\bar{w}\gets w$, where $Q(s,a;\bar{w})$ has the same architecture as the online network.
                
                \item[3.] \textbf{Initialize the replay buffer $\mathcal{D}$:} 
                A FIFO buffer storing transition tuples $(s,a,r,s',\mathrm{done})$ with capacity $N_{\mathrm{pool}}$.
                
                \item[4.] \textbf{Initialize the quantum sensing environment:} 
                The system is governed by a controlled parameter-dependent Hamiltonian 
                $H(\theta;\mathbf{u}_t)$, where $\theta$ is the unknown parameter and 
                $\mathbf{u}_t$ denotes the pulse-control parameters.
                
                \item[5.] \textbf{Construct the discrete action set $\mathcal{A}$:} 
                Discretize the experimentally tunable pulse parameters, such as phase, detuning, amplitude, 
                or pulse duration, to form a finite set of pulse-control actions.
            \end{itemize}
            
            \vspace{5pt}
            \textbf{Phase I: Replay-buffer pre-filling}
            
            \begin{itemize}
                \item[1.] \textbf{Reset the quantum environment:} 
                Prepare the nominal state $|\psi(\theta)\rangle$ and the perturbed state 
                $|\psi(\theta+d\theta)\rangle$.
                
                \item[2.] \textbf{Construct the initial RL state $s_0$:} 
                Concatenate the real and imaginary parts of the two quantum states. 
                For a $d$-dimensional Hilbert space, $s_0$ is a $4d$-dimensional real vector.
                
                \item[3.] \textbf{Select a fully random action:} 
                Set $\varepsilon=1$ and sample a pulse-control action $a_t$ uniformly from $\mathcal{A}$.
                
                \item[4.] \textbf{Execute the action and evolve the system:} 
                Propagate both $|\psi(\theta)\rangle$ and $|\psi(\theta+d\theta)\rangle$ under 
                the controlled Hamiltonian $H(\theta;\mathbf{u}_t)$.
                
                \item[5.] \textbf{Compute the reward:}
                \[
                r_t =
                \begin{cases}
                	\displaystyle
                	\log\left(\frac{F_C(t)}{t^2}\right),
                	& \text{Case A},\\[6pt]
                	\displaystyle
                	\Delta\log\left[1+F_C(t)\right],
                	& \text{Case B}.
                \end{cases}
                \]
                where $\mathcal{F}_C(t)$ denotes the classical Fisher information.
                
                \item[6.] \textbf{Store the transition:} 
                Add $(s_t,a_t,r_t,s_{t+1},\mathrm{done})$ to the replay buffer $\mathcal{D}$.
            \end{itemize}

            \vspace{5pt}
            \textbf{Phase II: Training}
            
            \begin{itemize}
                \item[1.] \textbf{Reset the environment at the beginning of each episode.}

                \item[2.] \textbf{Update the exploration rate:}
                \[
                    \varepsilon_{\mathrm{ep}}
                    =
                    \varepsilon_N
                    +
                    (\varepsilon_1-\varepsilon_N)
                    \exp\left(-\frac{8e}{N_{\mathrm{eps}}}\right),
                \]
                where $e$ is the current episode index.
                
                \item[3.] \textbf{Select an action using the $\varepsilon$-greedy policy:}
                \[
                    a_t =
                    \begin{cases}
                        \text{a random action from } \mathcal{A}, 
                        & \text{with probability } \varepsilon_{\mathrm{ep}},\\
                        \arg\max_{a\in\mathcal{A}} Q(s_t,a;w),
                        & \text{with probability } 1-\varepsilon_{\mathrm{ep}}.
                    \end{cases}
                \]
                
                \item[4.] \textbf{Execute the action and store the transition:} 
                Evolve the quantum system, compute $r_t$, obtain $s_{t+1}$, and store 
                $(s_t,a_t,r_t,s_{t+1},\mathrm{done})$ in $\mathcal{D}$.
                
                \item[5.] \textbf{Sample a mini-batch from the replay buffer:} 
                Sample a mini-batch uniformly at random from the replay buffer. If
                $|\mathcal{D}|\ge n_{\mathrm{batch}}$, draw
                $n_{\mathrm{batch}}$ transitions using uniform random sampling.
                
                \item[6.] \textbf{Compute the Double-DQN target:}
                \[
                    a^\ast =
                    \arg\max_{a'\in\mathcal{A}}
                    Q(s_{t+1},a';w),
                \]
                \[
                    y_t =
                    \begin{cases}
                        r_t, 
                        & \text{if } s_{t+1} \text{ is terminal},\\
                        r_t+\gamma Q(s_{t+1},a^\ast;\bar{w}),
                        & \text{otherwise}.
                    \end{cases}
                \]
                
                \item[7.] \textbf{Compute the loss:}
                \[\mathcal{L}(w)
                =
                \frac{1}{n_{\mathrm{batch}}}
                \sum_{(s,a,r,s',\mathrm{done})}
                \left[
                y_t-Q(s,a;w)
                \right]^2.
                \]
                
                \item[8.] \textbf{Update the online network:} 
                Update $w$ by gradient descent using the loss $\mathcal{L}(w)$.
                
                \item[9.] \textbf{Update the target network at the end of each episode:}
                \[
                \bar{w}\leftarrow w.
                \]
            \end{itemize}
            
            \vspace{2pt}
            \label{Table I}
        \end{minipage}%
    }
\end{table*}

\subsection{Training Algorithm}
To break the temporal correlations in the training data and improve sample efficiency, we employ an experience replay mechanism. During the interaction between the agent and the quantum environment, each transition tuple
\begin{equation}
	(s_t,a_t,r_t,s_{t+1},\mathrm{done})
\end{equation}
is stored in a circular replay buffer. During training, mini-batches of transitions are randomly sampled from the buffer to update the Q-network. The replay buffer is sampled uniformly at random during training. Mini-batches of transitions are drawn from the buffer once it contains enough samples for an update. The buffer is initialized with 80
transitions and has a maximum capacity of 1000 transitions, with the
oldest transitions discarded when the capacity is reached.

To improve training stability, we adopt a target-network architecture consisting of two neural networks with identical architectures: an online network parameterized by $w$ and a target network parameterized by $\bar{w}$. The input dimension is $4d$, corresponding to the real and imaginary components of the two parameter-dependent state vectors defined above. For Case A, $d=3$, and the input dimension is therefore 12. The network contains three hidden layers with 32, 64, and 64 neurons, respectively, with a sigmoid activation function applied after each hidden layer. The output layer is linear and has dimension $|\mathcal{A}|$, with one Q value assigned to each discrete control action. Accordingly, for Case A, the output dimension is 15 under phase-only control and 12 under joint phase--detuning control. For Case B, $d=2$, and the input dimension is therefore 8. The same network architecture is used. The output layer is linear and has dimension $|\mathcal{A}|$, with one Q value assigned to each available control action. The online network is used to select actions and evaluate the current Q values, while the target network is utilized to construct the bootstrapped target values. The target network is initialized with the online-network parameters and is updated after each completed training episode according to the
hard-update rule
\begin{equation}
	\bar{w}\leftarrow w.
\end{equation} This separation between action-value estimation and target construction mitigates the instability caused by rapidly moving targets in Q-learning.

Furthermore, we adopt the double DQN strategy to reduce the overestimation bias commonly caused by the maximization operation in standard Q-learning. Specifically, the online network is used to select the optimal next action,
\begin{equation}
    a_{t+1}^{\ast}
    =
    \arg\max_{a'\in\mathcal{A}}
    Q(s_{t+1},a';w),
\end{equation}
whereas the target network is employed to evaluate the value of this selected action. The target Q-value is therefore given by
\begin{equation}
    y_t
    =
    r_t
    +
    \gamma
    Q(s_{t+1},a_{t+1}^{\ast};\bar{w}),
\end{equation}
where $\gamma$ is the discount factor. The online network is optimized by minimizing the TD loss,
\begin{equation}
 \mathcal{L}(w)
 =
 \mathbb{E}_{(s_t,a_t,r_t,s_{t+1})}
 \left[
 \left(
 y_t-Q(s_t,a_t;w)
 \right)^2
 \right],
\end{equation}

The exploration strategy is based on the $\varepsilon$-greedy policy. The exploration rate $\varepsilon$ is initialized to unity and exponentially decays toward a prescribed minimum value during training. The agent selects a pulse-control action uniformly at random from the discretized action set $\mathcal{A}$ during exploration, whereas it selects the action with the largest estimated Q-value during exploitation. This strategy balances broad exploration of the pulse-control landscape with focused exploitation of the learned policy.

In this work, we employ a DQN-based training algorithm for the discretized pulse-control problem. The detailed procedure is summarized in Table~\ref{Table I}.
Unless otherwise specified, the numerical simulation parameters and reinforcement-learning hyperparameters are summarized in Table~\ref{Table II} of Appendix A. Based on these hyperparameters, we train the DRL agent to learn optimal pulse sequences for both the Hermitian three-level leakage system (Case A) and the non-Hermitian two-level system (Case B). The resulting CFI values are evaluated and compared under the corresponding physical and measurement models in the following section.

{\section{DRL-Based Control for Leakage Sensing in Two Model Systems}
\label{sec4}

\subsection{Case A: Hermitian three-Level System}
\label{sec:caseA}
We first apply the proposed DRL-based pulse-control framework to a typical $\Lambda$-type Hermitian three-level system. The Hilbert space is spanned by three basis states $\{|0\rangle,|1\rangle,|2\rangle\}$. In the present leakage-sensing setting, the states $|0\rangle$ and $|1\rangle$ define the computational subspace, while $|2\rangle$ represents the leakage state outside the computational subspace. The sensing task is to estimate the leakage-related parameter that governs the transition from the computational subspace to the leakage space.

In this case, the general controlled Hamiltonian introduced above can be specified as
\begin{equation}
    H(\eta;\mathbf{u}_t)
    =
    \frac{1}{2}
    \begin{pmatrix}
        0 & \Omega_p e^{i\phi_t} & 0\\
        \Omega_p e^{-i\phi_t} & 2\Delta_t & \Omega_s \eta\\
        0 & \Omega_s \eta & 0
    \end{pmatrix},
    \label{eq7}
\end{equation}
where $\eta$ is the leakage parameter to be estimated and
\begin{equation}
    \mathbf{u}_t=(\phi_t,\Delta_t)
\end{equation}
denotes the pulse-control parameters at the $t$-th control step. Here, $\Omega_p$ and $\Omega_s$ are the Rabi frequencies driving the transitions $|0\rangle \leftrightarrow |1\rangle$ and $|1\rangle \leftrightarrow |2\rangle$, respectively. The phase $\phi_t$ controls the complex phase of the pulse driving the $|0\rangle \leftrightarrow |1\rangle$ transition, while $\Delta_t=\omega_0-\omega_t$ is the detuning between the laser carrier frequency $\omega_t$ and the Bohr transition frequency $\omega_0$. The unknown parameter $\eta$ modulates the coupling strength between the computational state $|1\rangle$ and the leakage state $|2\rangle$. Therefore, estimating $\eta$ is equivalent to characterizing the strength of leakage from the computational subspace into the leakage space. For Case A, $\Delta t=1$ is fixed in normalized units. Therefore, the quadratic reference scaling $\mathcal{F}\propto T^2$ can be expressed as $\mathcal{F}\propto N^2$. This quadratic dependence is used only as a reference benchmark for the Hermitian three-level model.

\begin{figure}
	\centering
	\includegraphics[width=0.9\linewidth]{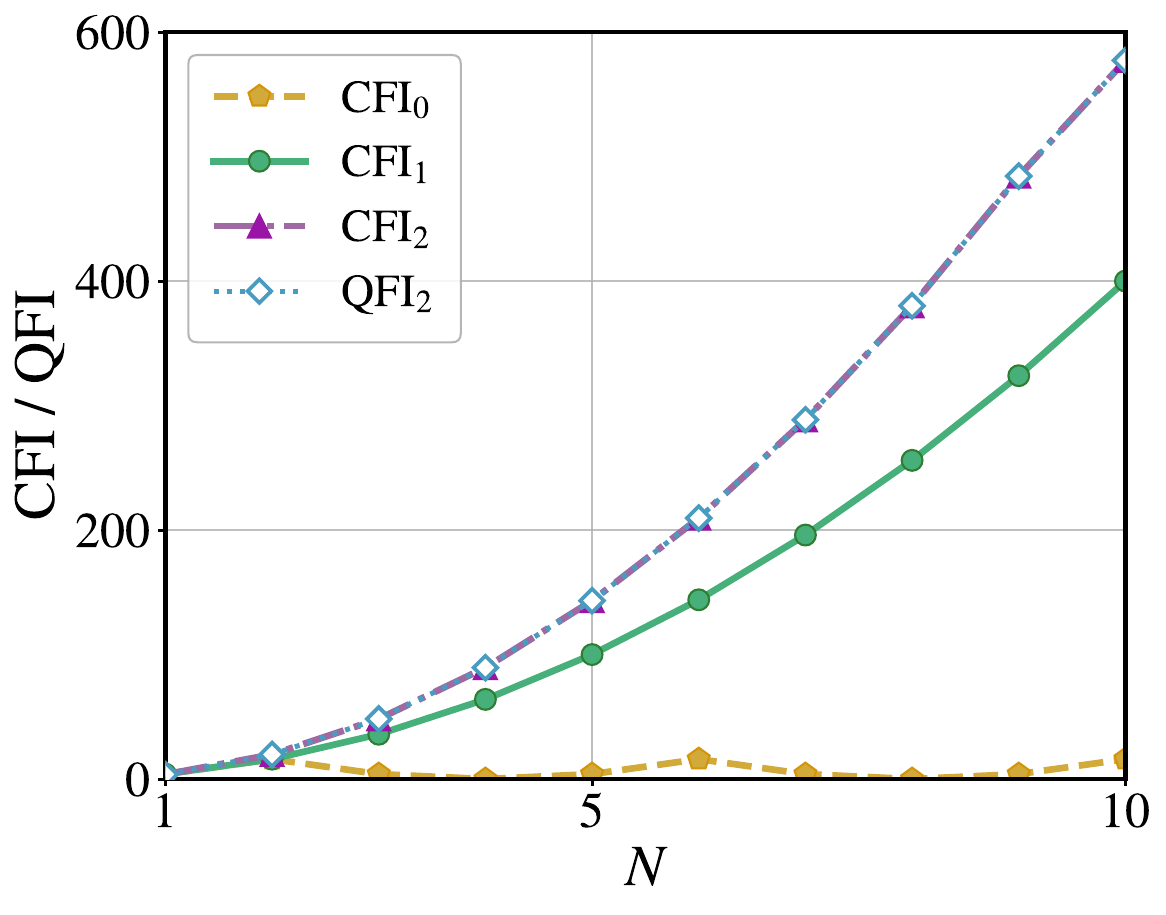}
	\caption{Classical and quantum Fisher information (CFI and QFI) as functions
		of the number of control pulses $N$ for the three-level leakage system.
		The horizontal axis denotes the pulse number $N$, and the vertical axis
		denotes the Fisher-information values. The yellow dashed, green solid,
		and purple dash-dot curves represent the fixed-resonant baseline
		$\mathrm{CFI}_0$, the phase-only control strategy $\mathrm{CFI}_1$, and
		the joint phase--detuning control strategy $\mathrm{CFI}_2$, respectively.
		The blue dotted curve with open diamond markers represents
		$\mathrm{QFI}_2$, evaluated for the same learned joint phase--detuning
		pulse sequence. The simulations are performed with $\hbar=1$,
		$\Delta t=1$, and $\Omega_p=\Omega_s=\pi$. All quantities are expressed
		in normalized units.}
	\label{fig1}
\end{figure}

We use a fixed-resonant baseline with
	$\phi_t=0$ and $\Delta_t=0$ for every control step, without adaptive
	pulse optimization. Its CFI is denoted by $\mathrm{CFI}_0$. In the one-parameter control setting, the agent optimizes only the pulse phase,
\begin{equation}
    \mathbf{u}_t=\phi_t,
\end{equation}
and the resulting Fisher information is denoted by $\mathrm{CFI}_1$. In the two-parameter collaborative control setting, the agent jointly optimizes the pulse phase and detuning as
\begin{equation}
    \mathbf{u}_t=(\phi_t,\Delta_t),
\end{equation}
and the corresponding Fisher information is denoted by $\mathrm{CFI}_2$. For a projective measurement in the energy-level basis, the CFI is
	given by
	\begin{equation}
		\mathcal{F}_C(\eta)
		=
		\sum_{i\in\{0,1,2\}}
		\frac{1}{p_i(\eta)}
		\left[
		\frac{\partial p_i(\eta)}{\partial\eta}
		\right]^2.
		\label{eq:cfi}
	\end{equation}
	The local sensitivity is therefore characterized by the population
	derivatives $\partial p_i/\partial\eta$. Larger derivative magnitudes can
	contribute more strongly to the CFI, provided that the corresponding
	probability weights remain finite and significant.
	
At the operating point $\eta_0=0$, direct substitution into
		Eq.~\eqref{eq:cfi} is not used because this point lies at the boundary
		of the physical parameter domain. During DQN training, the population derivatives at the boundary are
		evaluated using the forward finite-difference rule
		\begin{equation}
			\left.
			\frac{\partial p_i}{\partial\eta}
			\right|_{\eta=\eta_0}
			\approx
			\frac{
				p_i(\eta_0+\delta\eta)-p_i(\eta_0)}
			{\delta\eta},
			\qquad
			\delta\eta=10^{-3}.
			\label{eq:training_forward_difference}
		\end{equation}
		This training-stage finite-difference quantity is not taken to be the
		final boundary CFI.
For the final evaluation, the boundary value is defined by the
right-sided limit
\begin{equation}
	\mathcal{F}_{C}^{+}(0)
	=
	\lim_{\eta\rightarrow0^{+}}\mathcal{F}_{C}(\eta).
\end{equation}
Numerically, the right-sided limit is approximated by evaluating the
CFI at the positive interior points
\[
\epsilon\in
\{10^{-2},3\times10^{-3},10^{-3},3\times10^{-4},10^{-4}\},
\qquad h=\epsilon/10.
\] At each $\eta=\epsilon$, the probability derivatives are evaluated using
\begin{equation}
	\left.
	\frac{\partial p_i}{\partial\eta}
	\right|_{\eta=\epsilon}
	\approx
	\frac{p_i(\epsilon+h)-p_i(\epsilon-h)}{2h}.
\end{equation}
The Fisher-information contribution is evaluated using
$p_i(\epsilon)$ in the denominator, and terms with
$p_i(\epsilon)\leq10^{-30}$ are omitted. The value at the smallest
tested offset, $\epsilon=10^{-4}$, is used as the numerical
approximation to the right-sided boundary CFI. For the representative joint phase--detuning sequence, the relative change in the CFI between the two smallest tested offsets, $\epsilon=3\times10^{-4}$ and $10^{-4}$, is approximately $5.4\times10^{-7}$, providing evidence that the right-sided-limit estimate is numerically stable. The learned pulse
sequence is kept unchanged during this post-training reevaluation. The boundary-aware treatment affects only the evaluation of the CFI and
does not alter the population trajectories generated by the specified
Hamiltonian and pulse sequences.
	\begin{figure}
		\centering
		\includegraphics[height=4cm]{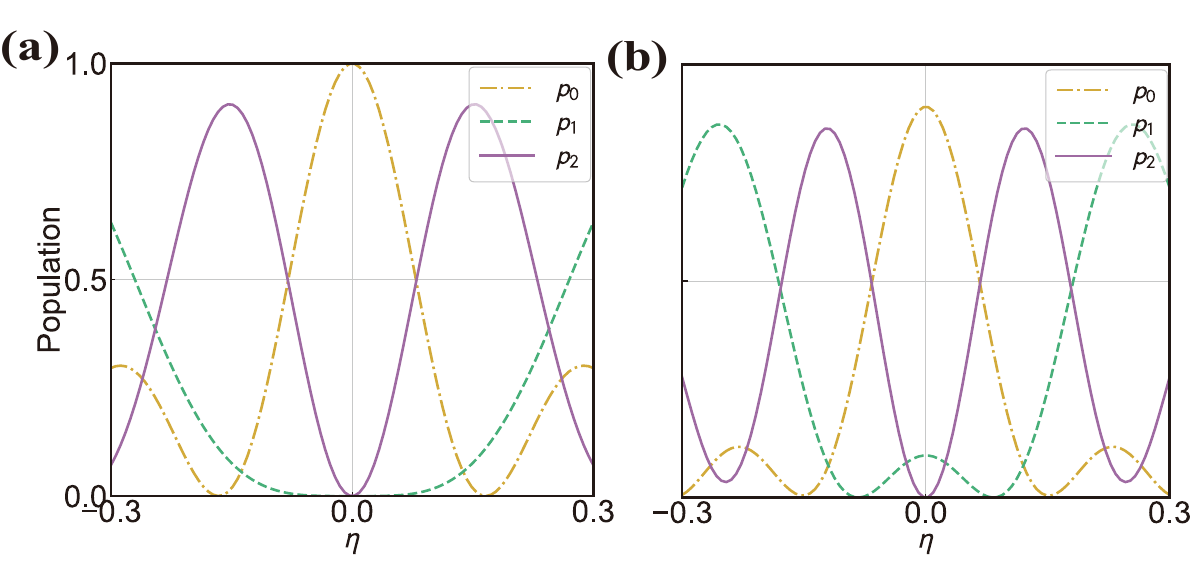}
		\caption{Population response of the three-level system under different pulse-control dimensions. The horizontal axis denotes the leakage parameter $\eta$, and the vertical axis denotes the population probability.
			Panel (a) shows phase-only control, whereas panel (b)
			shows joint phase--detuning control. The yellow dash-dot, green dashed,
			and purple solid curves represent the populations
			$p_0(\eta)$, $p_1(\eta)$, and $p_2(\eta)$, respectively. The simulations are performed with $\hbar = 1,\; \Delta t = 1,\; \Omega_p = \Omega_s = \pi$. All quantities are expressed in normalized units.}
		\label{fig2}
	\end{figure}
For the same learned joint phase--detuning pulse sequence, we also evaluate the corresponding quantum Fisher information, denoted by $\mathrm{QFI}_2$. For each pulse number $N$, $\mathrm{QFI}_2(N)$ is calculated from the parameter-dependent quantum state obtained after propagating the first $N$ pulses, using the QFI definition introduced above. Since $\eta=0$ is a boundary point in Case A, the final QFI is evaluated using the regular right-sided limit $\eta\rightarrow0^{+}$. The pulse sequence is kept unchanged during this evaluation, and no additional control optimization is performed. The CFI and $\mathrm{QFI}_2$ trends shown in Fig.~\ref{fig1}	demonstrate that, as the number of control pulses increases, both  optimized protocols achieve substantially larger CFI values than the fixed-resonant baseline, with the joint phase--detuning protocol consistently outperforming the phase-only protocol. The $\mathrm{QFI}_2$ curve closely tracks $\mathrm{CFI}_2$ over the plotted pulse range. This near coincidence indicates that the specified projective measurement extracts most of the information available from the evolved state under this protocol.

The performance gap between $\mathrm{CFI}_1$ and $\mathrm{CFI}_2$ can be understood from the distinct roles of the pulse-control parameters in Eq.~\eqref{eq7}. The phase $\phi_t$ mainly controls the relative phase of the $|0\rangle\leftrightarrow |1\rangle$ transition within the computational subspace. By contrast, the detuning $\Delta_t$ modifies the diagonal energy of the intermediate state $|1\rangle$ and shifts the effective resonance condition for population transfer toward the leakage state $|2\rangle$. The two-parameter collaborative strategy can therefore control both the phase structure inside the computational subspace and the resonance condition associated with leakage. This additional flexibility allows the system to access more favorable quantum trajectories, in which the state remains highly sensitive to variations in $\eta$.

To further clarify the advantage of $\mathrm{CFI}_2$ over $\mathrm{CFI}_1$, we analyze the sensing dynamics from the perspective of parameter-dependent population responses. In the present leakage-sensing problem, the occupation probability of the leakage state,
\begin{equation}
    p_2(\eta)=|\langle 2|\psi(\eta)\rangle|^2,
\end{equation}
serves as a direct indicator of the leakage strength. Since $\eta$ modulates the coupling between $|1\rangle$ and $|2\rangle$, the response of $p_2(\eta)$ to changes in $\eta$ provides a sensitive probe of leakage.

Figure~\ref{fig2} compares the level populations
$(p_0,p_1,p_2)$ as functions of the leakage parameter $\eta$ under the
two control strategies. Under phase-only control, as shown in
Fig.~\ref{fig2}(a), the leakage-state population $p_2(\eta)$ exhibits
smooth, approximately oscillatory variations. Under joint
phase--detuning control, as shown in Fig.~\ref{fig2}(b), the population
curves exhibit sharper and more localized variations with $\eta$. These
results show that the joint-control sequence produces a stronger
population response to variations in the leakage parameter over the
range considered.
\begin{figure}
	\centering
	\includegraphics[height=3.7cm]{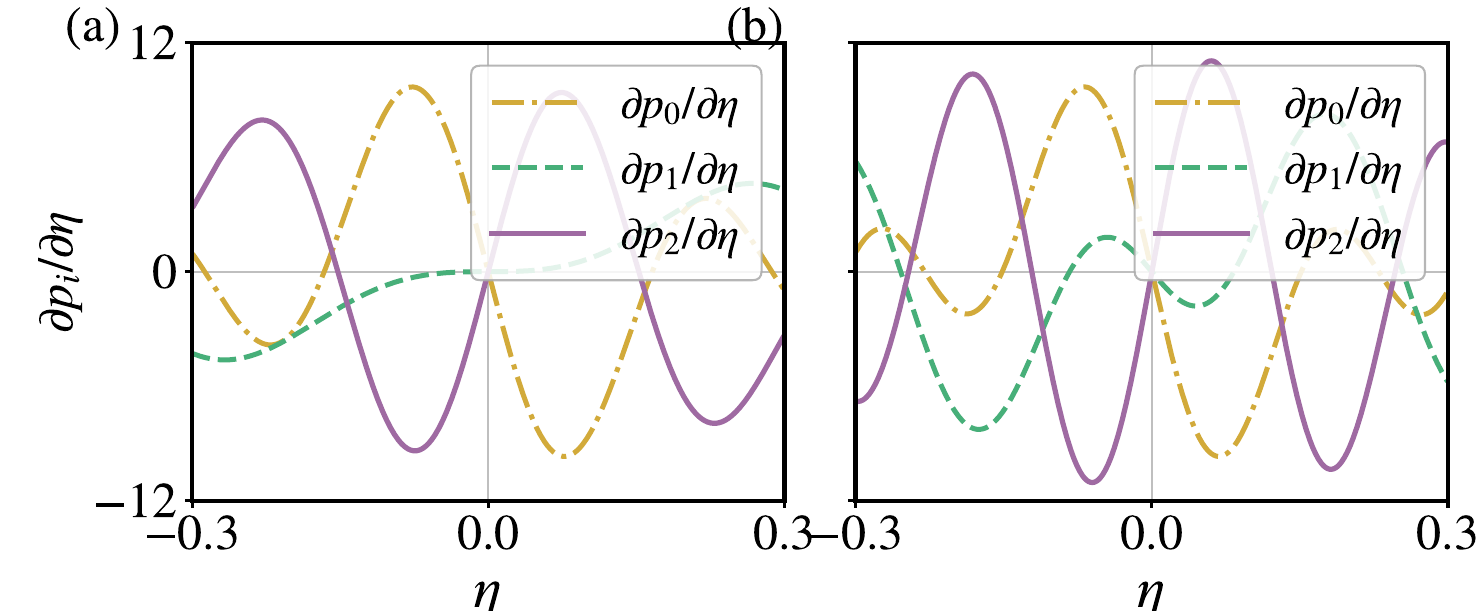}
	\caption{Partial derivatives of the level populations with respect to the
		leakage parameter $\eta$, $\partial p_i/\partial\eta$, as functions of
		$\eta$ for the three-level leakage system. The horizontal axis denotes
		the leakage parameter $\eta$, and the vertical axis denotes the
		population derivative $\partial p_i/\partial\eta$. Panel (a) shows
		phase-only control, whereas panel (b) shows joint phase--detuning
		control. The yellow dash-dot, green dashed, and purple solid curves
		represent $\partial p_0/\partial\eta$,
		$\partial p_1/\partial\eta$, and
		$\partial p_2/\partial\eta$, respectively. The simulations are performed with $\hbar=1$, $\Delta t=1$, and $\Omega_p=\Omega_s=\pi$. All quantities are expressed in normalized units.}
	\label{fig3}
\end{figure}

Compared with the one-parameter control result in Fig.~\ref{fig3}(a), the two-parameter collaborative control in Fig.~\ref{fig3}(b) produces population-derivative peaks that are higher and more concentrated. These sharper derivative features explain the enhanced CFI observed in Fig.~\ref{fig1}. Physically, the additional detuning control allows the agent to tune the system closer to parameter-sensitive resonance conditions, while the phase control adjusts the coherent transition pathway. Their combined effect enlarges the set of accessible quantum trajectories and enables the agent to concentrate sensitivity into regions where the measurement probabilities respond strongly to changes in $\eta$.

Overall, under the specified Hamiltonian, control parameterization, and
	projective measurement model, the DQN-discovered joint phase--detuning pulse sequence yields a CFI that closely approaches the corresponding QFI and is consistent with
	the quadratic reference scaling over the pulse-number range considered. These results show that pulse-level reinforcement learning can identify control sequences that extract most of the available information from the evolved state and enhance measurement-model-specific leakage sensitivity. We emphasize that the leakage considered in Case A refers to population transfer from the computational subspace ($|0\rangle$,$|1\rangle$) to the leakage state ($|2\rangle$), which is explicitly retained in the Hilbert space. Thus, the Hamiltonian remains Hermitian, and the overall three-level evolution is unitary and trace-preserving. In this sense, Case A describes computational-subspace leakage rather than irreversible population loss from the full system.

\begin{figure}
	\centering
	\includegraphics[width=0.9\linewidth]{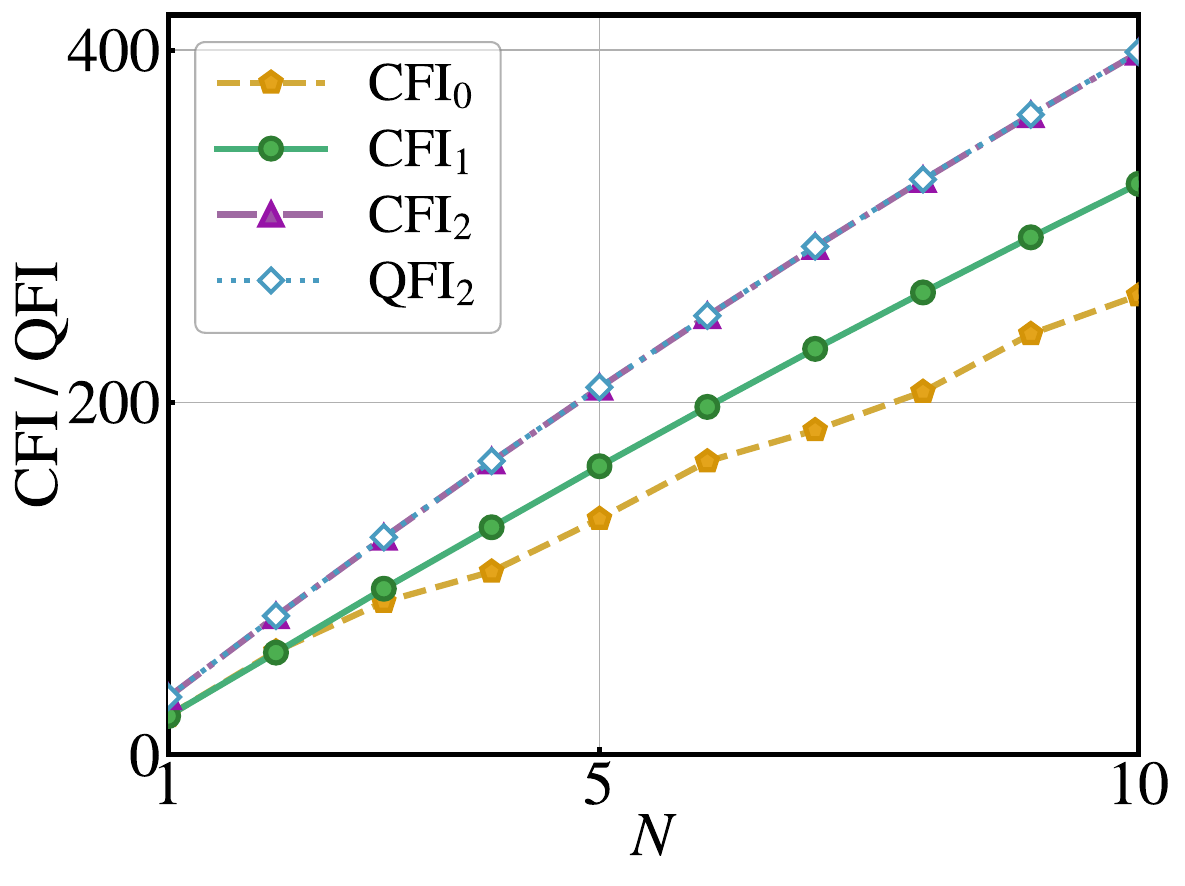}
	\caption{
			Classical and quantum Fisher information for the effective
			non-Hermitian two-level model as functions of the number of control
			pulses $N$. The yellow dashed, green solid, and purple dash-dot curves
			represent the fixed-resonant baseline $\mathrm{CFI}_0$, the phase-only
			control strategy $\mathrm{CFI}_1$, and the joint phase--detuning control
			strategy $\mathrm{CFI}_2$, respectively. The blue dotted curve with open
			diamond markers represents $\mathrm{QFI}_2$. The simulations are
			performed with $\hbar=1$, $\Delta t=1$, and $\Omega=\pi$. All quantities
			are expressed in normalized units.
	}
	\label{fig4}
\end{figure}

\subsection{Case B: Non-Hermitian Two-Level System}

We next consider a non-Hermitian two-level system as an effective model for sensing dissipation-induced attenuation. Compared with the three-level leakage model in Case A, the leakage state is not explicitly included in Case B. Instead, its effect is represented by the non-Hermitian attenuation term $-i\eta$ in the Hamiltonian acting on the two-level subspace spanned by $|0\rangle$ and $|1\rangle$. In this setting, the parameter to be estimated is the intrinsic attenuation rate $\eta$. The operating point is set to
$\eta_0=0.025$.

In the basis $\{|0\rangle,|1\rangle\}$, the controlled Hamiltonian is given by
\begin{equation}
	H(\eta;\mathbf{u}_t)
	=
	\frac{1}{\sqrt{2}}
	\begin{pmatrix}
		2\Delta_t & \Omega e^{i\phi_t}\\
		\Omega e^{-i\phi_t} & -i\eta
	\end{pmatrix},
	\label{eq:nonhermitian_H}
\end{equation}
where $\Omega$ is the Rabi frequency, $\phi_t$ is the driving phase, and $\Delta_t$ is the detuning. The imaginary term $-i\eta$ describes attenuation of the state $|1\rangle$, thereby encoding the attenuation-rate information into the non-unitary dynamics. 
For the non-Hermitian two-level model, the state evolution is generally
non-unitary because of the effective attenuation term. Let
$|\tilde{\psi}(\eta)\rangle$ denote the resulting, generally
unnormalized, two-level state. The probabilities of detecting the system
in the two retained states are defined as
\begin{equation}
	p_0(\eta)=|\langle 0|\tilde{\psi}(\eta)\rangle|^2,
	\qquad
	p_1(\eta)=|\langle 1|\tilde{\psi}(\eta)\rangle|^2.
\end{equation}
Because the observation model contains the two detected outcomes and
the complementary loss outcome, the probabilities satisfy
\begin{equation}
	p_0(\eta)+p_1(\eta)+p_{\mathrm{loss}}(\eta)=1.
\end{equation}
Operationally, we assume a counting experiment in which the total
number of interrogation trials is known and each trial yields exactly
one of three mutually exclusive outcomes: detection in $|0\rangle$,
detection in $|1\rangle$, or a loss/no-click event. The probability of
the third outcome is
$p_{\mathrm{loss}}(\eta)=1-p_0(\eta)-p_1(\eta)$. In this idealized
observation model, detector inefficiency and other unmodeled loss
mechanisms are not included explicitly; the loss/no-click outcome
represents the complementary probability associated with the effective
two-level description. The CFI is evaluated from this complete normalized observation model,
\begin{equation}
	\mathcal{F}_C(\eta)
	=
	\sum_{x\in\{0,1,\mathrm{loss}\}}
	\frac{1}{p_x(\eta)}
	\left[
	\frac{\partial p_x(\eta)}{\partial\eta}
	\right]^2.
\end{equation}
}\begin{figure}[t]
\centering
\includegraphics[width=0.9\linewidth]{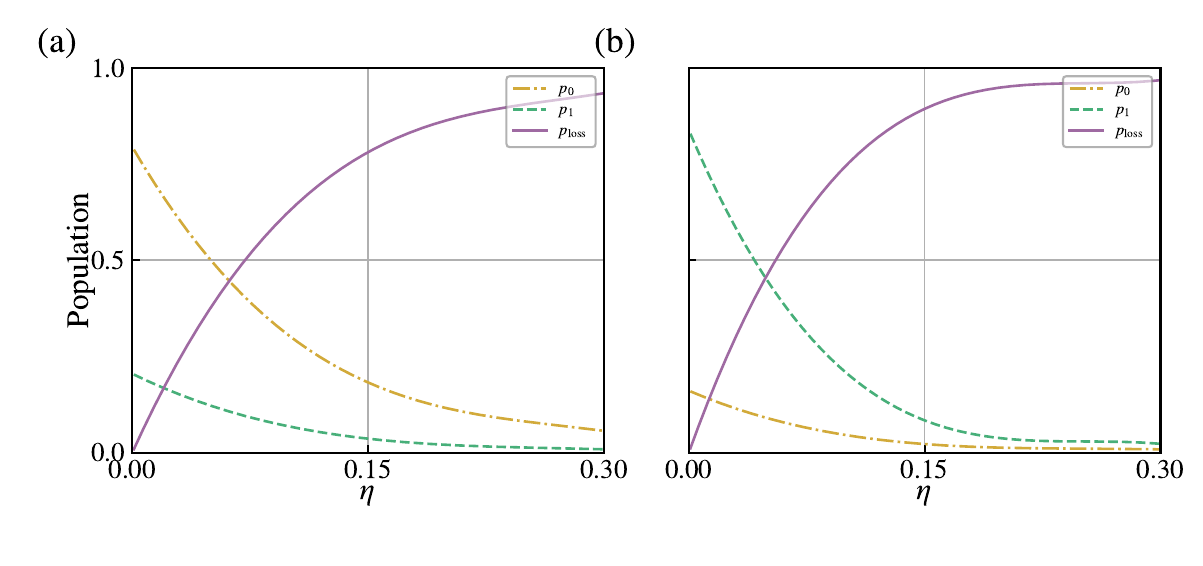}
\caption{Normalized observation probabilities versus the attenuation parameter $\eta$ for the effective non-Hermitian two-level system. Panel (a) shows phase-only control, whereas panel (b) shows joint phase--detuning control. The yellow dash-dot, green dashed, and purple solid curves denote $p_0$, $p_1$, and $p_{\mathrm{loss}}$, respectively, where $p_{\mathrm{loss}}$ denotes the complementary outcome associated with loss from the effective two-level subspace. They satisfy $p_0+p_1+p_{\mathrm{loss}}=1$ and jointly define the observation model used for the CFI. Here, $\hbar=1$, $\Delta t=1$, and $\Omega=\pi$.}
\label{fig5}
\end{figure}
Although $|\tilde{\psi}(\eta)\rangle$ is generally unnormalized, it can
	be embedded into a normalized observation state by explicitly including
	an orthogonal loss/no-click flag $|L\rangle$. Let
	\begin{equation}
	s(\eta)=\langle\tilde{\psi}(\eta)|\tilde{\psi}(\eta)\rangle
   \end{equation}
	denote the survival probability of the retained two-level branch. By the
	definitions above, $s(\eta)=p_0(\eta)+p_1(\eta)$, and the previously
	defined $p_{\mathrm{loss}}(\eta)$ therefore corresponds to the
	complementary loss outcome. The normalized observation state is	\begin{equation}
	\rho_\eta=
	|\tilde{\psi}(\eta)\rangle\langle\tilde{\psi}(\eta)|
	+p_{\mathrm{loss}}(\eta)|L\rangle\langle L|,\end{equation}
	where $|L\rangle$ is orthogonal to the retained two-level subspace.
	The state $\rho_\eta$ is normalized and represents the complete
	three-outcome observation model.
	For the retained two-level branch, we define	
	\begin{equation}
	|\varphi(\eta)\rangle
	=
	\frac{|\tilde{\psi}(\eta)\rangle}{\sqrt{s(\eta)}}.\end{equation}
	The QFI of the complete observation state can then be written as
	\begin{equation}
	\mathcal{F}_{Q,\mathrm{comp}}
	=
	s(\eta)\mathcal{F}_{Q}[|\varphi(\eta)\rangle]
	+
	\frac{[\partial_\eta s(\eta)]^2}{s(\eta)},
\end{equation}
	and
	\begin{equation}
	\mathcal{F}_{Q,\mathrm{loss}}
	=
	\frac{[\partial_\eta p_{\mathrm{loss}}(\eta)]^2}
	{p_{\mathrm{loss}}(\eta)}.
\end{equation}
	Thus, the total QFI used for the Case-B joint phase--detuning control protocol is
	\begin{equation}
	\mathcal{F}_{Q,\mathrm{total}}
	=
	\mathcal{F}_{Q,\mathrm{comp}}
	+
	\mathcal{F}_{Q,\mathrm{loss}}.
\end{equation}
 This construction includes the information carried by both the survival probability and the loss outcome. The resulting QFI was independently verified using the symmetric-logarithmic-derivative expression for the complete three-dimensional density matrix. This treatment accounts explicitly for the loss outcome rather than	conditioning on detection within the retained two-level subspace. The	CFI curves in Fig.~\ref{fig4} show that the fixed-resonant baseline	remains below the optimized control protocols under the same complete observation model. Phase-only control improves the information accumulation, while joint phase--detuning control provides the largest measurement-model-specific CFI. For the joint phase--detuning pulse sequence discovered by the DQN, $\mathrm{CFI}_2$ closely approaches the corresponding $\mathrm{QFI}_2$. Moreover, $\mathrm{CFI}_2$ increases approximately linearly with the pulse number $N$ over the range considered.

\begin{figure}
	\centering
	\includegraphics[width=1.0\linewidth]{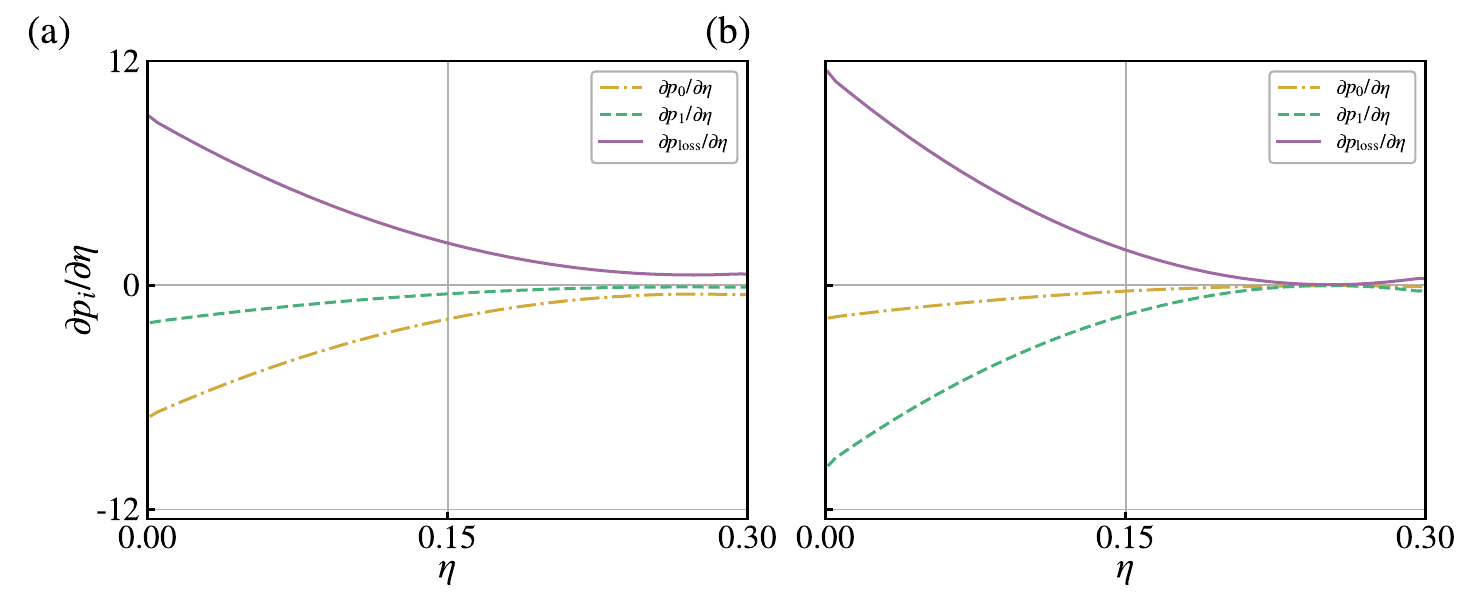}
\caption{Derivatives of the normalized observation probabilities with respect to the attenuation parameter $\eta$ for the effective non-Hermitian two-level system. Panel (a) shows phase-only control, whereas panel (b) shows joint phase--detuning control. The curves represent $\partial p_0/\partial\eta$, $\partial p_1/\partial\eta$, and $\partial p_{\mathrm{loss}}/\partial\eta$. Because $p_0+p_1+p_{\mathrm{loss}}=1$, the three derivatives sum to zero up to numerical precision.}
	\label{fig6}
\end{figure}
The advantage of the two-parameter strategy can be traced to the different physical roles of the two pulse parameters. The phase $\phi_t$ controls the coherent transition pathway between $|0\rangle$ and $|1\rangle$, whereas the detuning $\Delta_t$ changes the energy bias and effective resonance condition. In the presence of non-Hermitian attenuation, this additional detuning control allows the agent to balance coherent population transfer against dissipative loss, thereby steering the system toward trajectories that remain more sensitive to $\eta$. This mechanism is reflected in the population responses shown in Fig.~\ref{fig5}. Under phase-only control, the observation probabilities $p_0(\eta)$, $p_1(\eta)$, and $p_{\mathrm{loss}}(\eta)$ exhibit relatively smooth variations, indicating that the accessible trajectories are constrained by the fixed detuning structure. By contrast, joint phase--detuning control produces stronger and more asymmetric population changes. Since the attenuation term acts directly on $|1\rangle$, the
parameter dependence of $p_1(\eta)$ provides a direct contribution to the estimation of the attenuation rate. The complementary probability $p_{\mathrm{loss}}(\eta)$ also contributes
to the CFI, while its inclusion completes the normalized observation
model. Because $p_0(\eta)+p_1(\eta)+p_{\mathrm{loss}}(\eta)=1$, the three derivatives
sum to zero up to numerical precision. Their combined contributions,
weighted according to the CFI expression above, determine the measurement sensitivity. The derivative profiles in Fig.~\ref{fig6} further confirm this interpretation.  Compared with phase-only control, the two-parameter strategy produces larger and more pronounced observation-probability derivatives $\partial p_i/\partial\eta$, including $\partial p_{\mathrm{loss}}/\partial\eta$, especially in the range $[0,0.3]$. These enhanced derivative features indicate that the learned control sequence strengthens the parameter response, which directly explains the higher CFI observed in Fig.~\ref{fig4}. The results also show that the agent is not merely amplifying the original response, but actively reshaping the sensitivity landscape through detuning-assisted control.

 \subsection{Benchmark Comparison}
To provide a controlled comparison, the proposed DQN framework was
   	compared with random search (RS), best-of-random search (BOR), CRAB,
   	and GRAPE. Within each physical model, all methods use the same initial
   	state, Hamiltonian, pulse duration, number of control pulses, admissible
   	control ranges, and measurement model.
   	\begin{figure}
   		\centering
   		\includegraphics[width=1.0\linewidth]{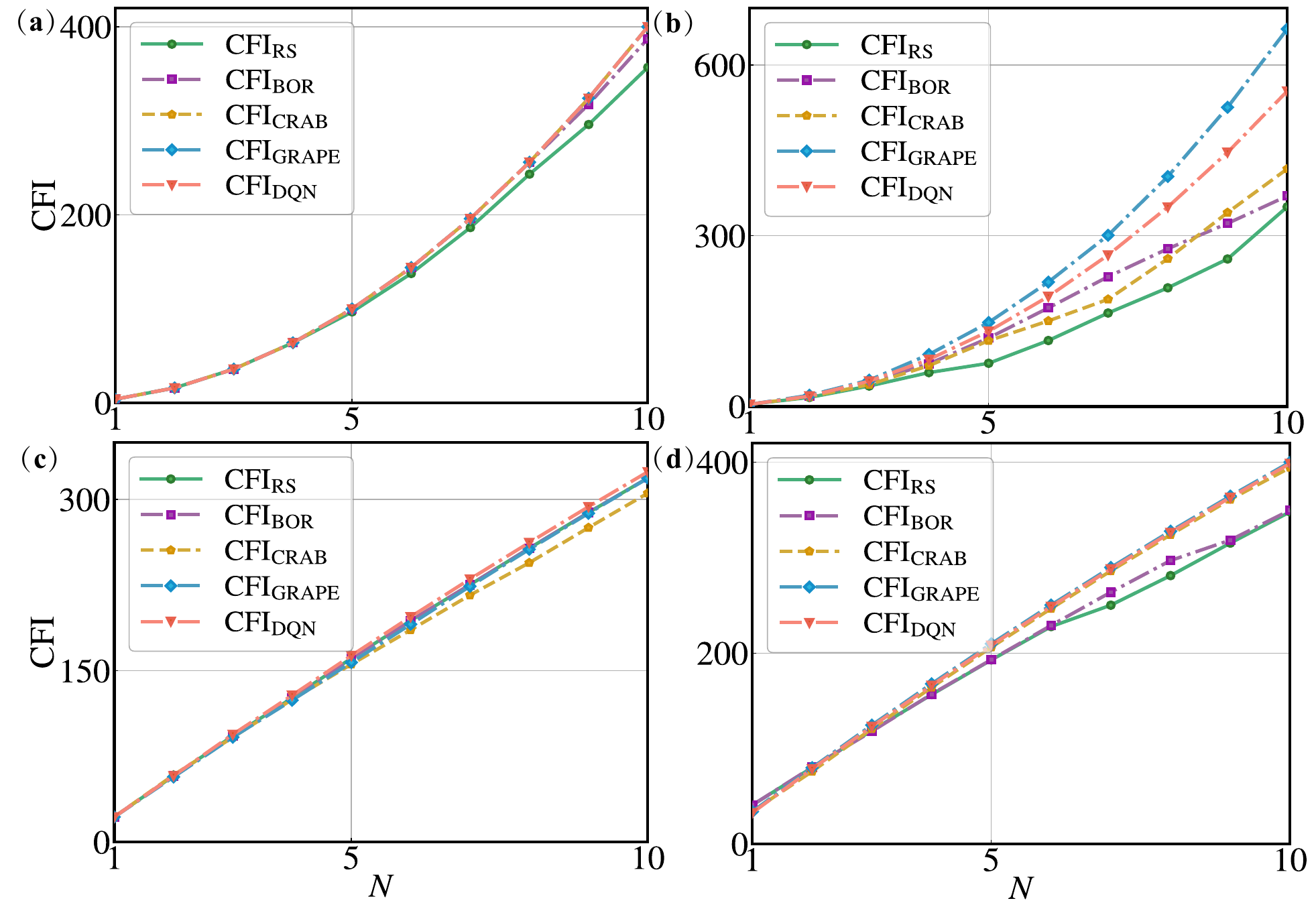}
   		\caption{
   				Comparison of random search (RS), best-of-random search (BOR), CRAB,
   				GRAPE, and the proposed DQN framework for the two physical models.
   				The horizontal axis denotes the number of control pulses $N$, and the
   				vertical axis denotes the CFI. Panels (a) and (b) show the Case-A
   				results for phase-only and joint phase--detuning control, respectively,
   				whereas panels (c) and (d) show the corresponding Case-B results. In all panels, GRAPE denotes a constrained control-optimization method operating over the prescribed control ranges. Within each physical model, all methods use the same Hamiltonian, pulse duration, control constraints, and measurement setting.
   		}
   		\label{fig7}
   	\end{figure}
   	
   For the Case-A results shown in Figs.~\ref{fig7}(a) and (b), RS
   		constructs pulse sequences by independently sampling actions from the
   		prescribed action space and evaluates 600 randomly generated sequences.
   		BOR generates 2000 complete candidate sequences and selects the sequence
   		with the largest final CFI. CRAB parameterizes the control sequence
   		using a truncated Fourier basis and optimizes the corresponding
   		coefficients with the Nelder--Mead method for at most 600 iterations.
   	The GRAPE comparator performs constrained control optimization over the prescribed control ranges. The phase is restricted to $[0,2\pi)$, while the detuning takes values in $\{-1,0,1\}$. The optimization starts from a randomly generated sequence and allows at most 600 update sweeps. For Case B, all final CFI values are evaluated from the complete normalized three-outcome observation model containing $p_0$, $p_1$,	and $p_{\mathrm{loss}}$. For the Case-B results shown in Figs.~\ref{fig7}(c) and (d), the	comparison is performed for the effective non-Hermitian two-leve model. RS constructs pulse sequences by independently sampling the allowed control actions and evaluates 600 randomly generated sequences. BOR generates 2000 complete candidate sequences and selects the sequence with the largest final CFI. CRAB parameterizes the control sequence using a truncated Fourier basis and optimizes the corresponding coefficients with the Nelder--Mead method for at most 600 iterations. For phase-only control, the phase is restricted to $[0,2\pi)$, while the detuning is fixed at zero. For joint phase--detuning control, the phase is restricted to $[0,2\pi)$, and the detuning takes values in $\{-3,-2,-1,0,1,2,3\}$. The GRAPE comparator starts from a randomly generated sequence and allows at most 600 update sweeps.

   The method-specific search budgets count different types of operations
   	and therefore do not represent equivalent computational costs.
   	Accordingly, Fig.~\ref{fig7} compares the achieved CFI values under the
   	stated optimization settings rather than the computational efficiency
   	of the different methods. The DQN training procedure and the associated reinforcement-learning
   	parameters are summarized in Table~\ref{Table II}. The benchmark results are summarized in Fig.~\ref{fig7}, which compares the CFI obtained by the five optimization methods as a function of the	number of control pulses $N$ for both control settings in Cases A and B. In Case A, DQN and GRAPE use their respective control
   	representations under the same physical model and admissible
   	control ranges. Their comparison evaluates the optimization procedures under the specified settings, without assuming identical discrete control spaces. In the representative Case-A benchmark runs, the DQN-generated
   sequences achieve competitive final CFI values relative to RS, BOR,
   CRAB, and GRAPE under the same physical model,
   admissible control ranges, and measurement model. At $N=10$, the
   boundary-aware CFI values obtained with DQN are approximately $400.0$
   for phase-only control and $577.2$ for joint phase--detuning control,
   as shown in Fig.~\ref{fig7}(a) and (b). For Case B, the DQN-generated sequences yield final CFI values of approximately $\mathrm{CFI}_1=324.2$ and $\mathrm{CFI}_2=398.7$ at $N=10$ for phase-only and joint phase--detuning control, respectively, as shown in panels (c) and (d) of Fig.~\ref{fig7}. These values are evaluated using the complete normalized observation model containing $p_0$, $p_1$, and $p_{\mathrm{loss}}$. Under the specified model and control settings, the larger value of $\mathrm{CFI}_2$ shows that joint phase--detuning control provides higher measurement-model-specific sensitivity than phase-only control.

   	Overall, under the specified physical models, observation models,
   		control constraints, and search budgets, the representative benchmark
   		results show that DQN achieves competitive CFI values in both cases,
   		without implying global optimality or a universal advantage in
   		computational efficiency.

 \subsection{Training Stability and Experimental Feasibility}
\begin{figure}
 	\centering
 	\includegraphics[width=1.0\linewidth]{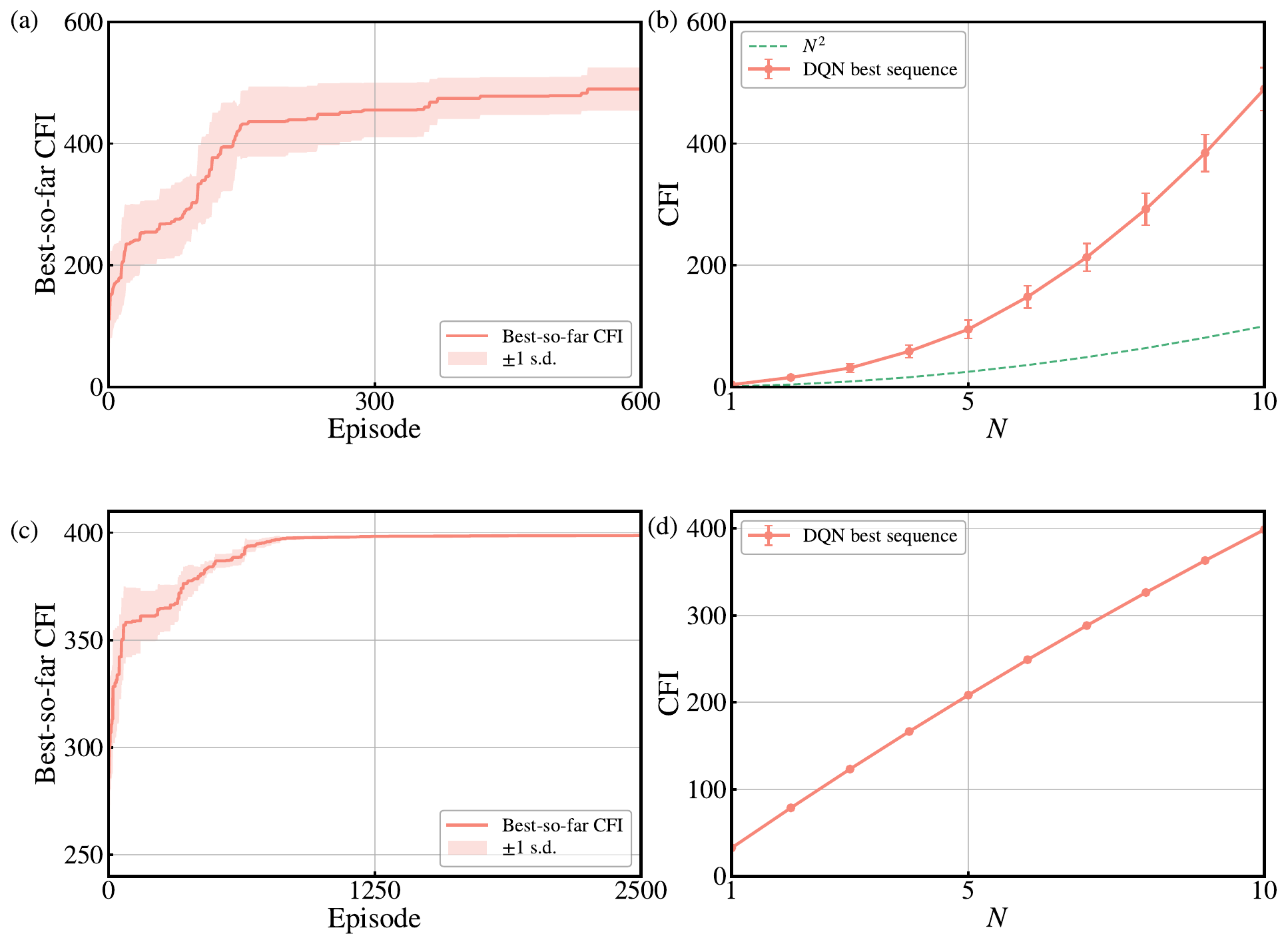}
 	\caption{
 				Reproducibility of the DQN-based control optimization for Case A and
 				Case B. Panels (a) and (b) show the Case-A results, whereas panels (c)
 				and (d) show the Case-B results. All panels report results for joint phase--detuning control. Panels (a) and (c) show
 				the mean best-so-far CFI during training, while panels (b) and (d) show
 				the mean CFI obtained from the best pulse sequence selected in each run as a function of the
 				cumulative pulse number $N$. Solid curves denote means over ten random
 				seeds; shaded regions and error bars represent one population standard
 				deviation.
 	}
 	\label{fig8}
 \end{figure}

The ten random seeds used for the training-stability analysis are 2016--2025. Unless otherwise specified, representative single-run results use the default seed 2024. Fig.~\ref{fig8} reports statistics aggregated over these ten independent runs. 
 For Case A, each training run consists of 600 episodes, whereas each
 Case-B training run consists of 2500 episodes.
 At the beginning of each run, the Python, NumPy, and PyTorch
 random-number generators are initialized with the corresponding seed. Within each physical model, all other Hamiltonian, pulse-control, reward, and network parameters are kept unchanged across the runs.
 For Case A, the boundary CFI is evaluated as
 $\mathcal{F}_C^{+}(0)=\lim_{\eta\rightarrow0^{+}}\mathcal{F}_C(\eta)$. For Case B, the CFI is evaluated from the complete normalized observation model containing the outcomes
 $p_0$, $p_1$, and $p_{\mathrm{loss}}$. During each episode, the final
 CFI of the generated pulse sequence and the best CFI observed up to that
 episode are recorded. After training, the pulse sequence associated
 with the best observed CFI is selected for each seed and independently
 reevaluated as a function of the cumulative pulse number $N$. The mean and population standard deviation over the ten independent runs
 are then used to generate all curves and error bars in
 Fig.~\ref{fig8}.

   The training stability of the DQN framework is illustrated in panels
   	(a) and (c) of Fig.~\ref{fig8} for Case A and Case B, respectively. These panels show the mean best-so-far CFI
   	as a function of the training episode. The solid curves denote the
   	means over ten independent random seeds, while the shaded regions
   	represent one population standard deviation. The Case-A curves contain
   	600 training episodes, whereas the Case-B curves contain 2500 training
   	episodes. In both physical models, the best-so-far CFI increases
   	rapidly during the early stage of training and subsequently stabilizes
   	after several hundred episodes, although the detailed convergence
   	behavior depends on the control setting and physical model.
  The run-to-run variation of the learned control protocols is further
  evaluated in panels (b) and (d) of Fig.~\ref{fig8} for Case A and
  Case B, respectively. These panels show the mean CFI
   	as a function of the cumulative pulse number $N$, together with error
   	bars representing one population standard deviation. The error bars quantify the run-to-run variation associated with different random initializations. For Case A, the mean CFI exhibits approximately quadratic growth with $N$ and follows the $N^2$ reference
   	trend over the investigated pulse range. At $N=10$, the ten-seed mean
   	CFI is approximately $489.6$, with a population standard deviation of
   	$35.3$. By contrast, the Case-B CFI increases approximately linearly
   	with $N$ over the investigated range. These results characterize the
   	statistical reproducibility and the distinct information-growth trends
   	of the learned joint phase--detuning protocols in the two models. For completeness, the representative learned pulse sequences obtained by the proposed
   	DQN framework for Case A and Case B are listed in Table~\ref{Table III} of
   	Appendix B.

\begin{figure}
\centering
\includegraphics[width=0.9\linewidth]{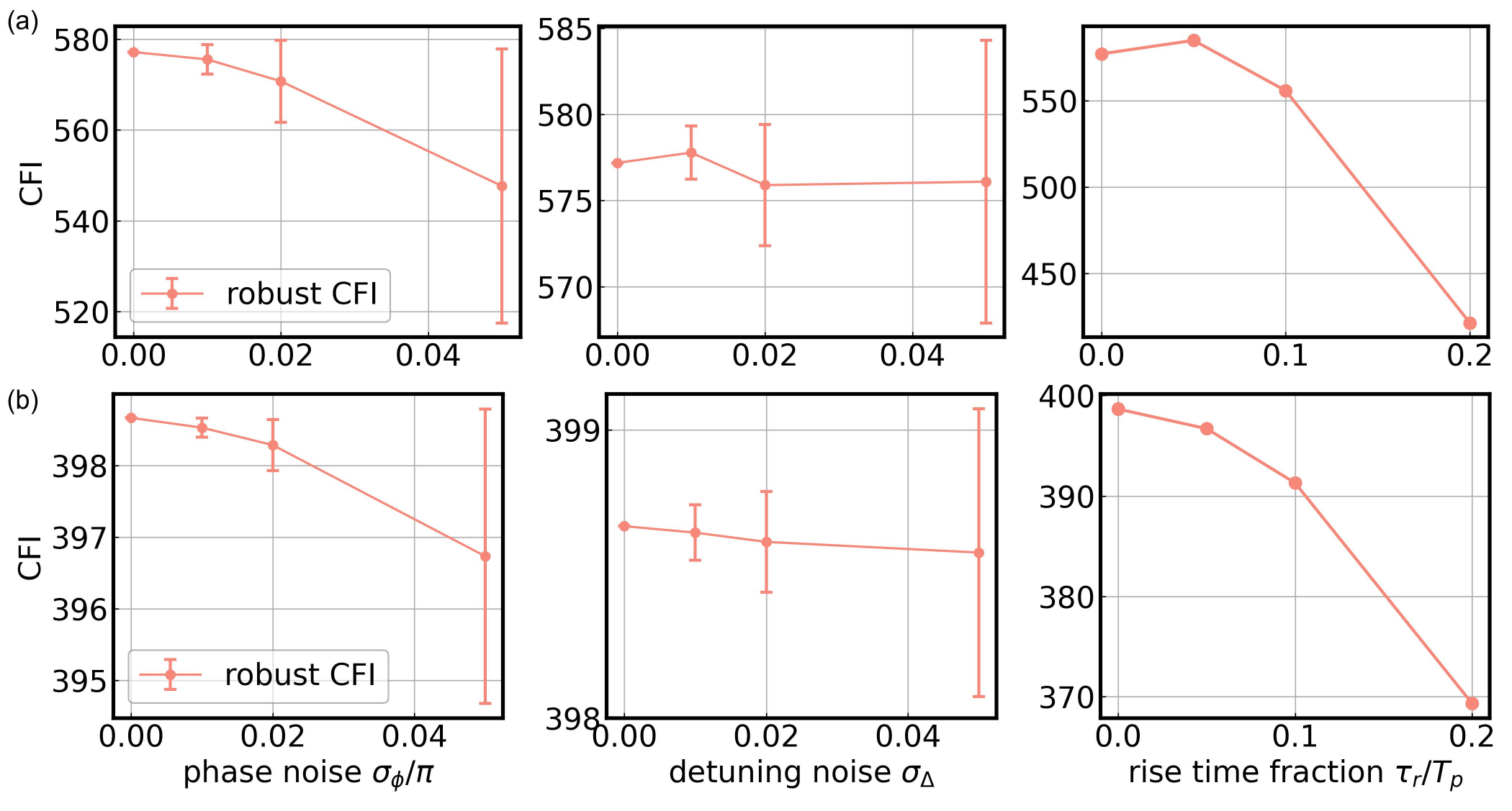}
\caption{
		Robustness of the learned joint phase--detuning control protocols under modeled control imperfections. Row (a) shows the results for the
		Hermitian three-level leakage model (Case A), whereas row (b) shows the
		results for the effective non-Hermitian two-level model (Case B). Within
		each row, the left, middle, and right panels show the CFI as a function
		of phase calibration noise, detuning calibration noise, and the
		normalized pulse rise-time fraction $\tau_r/T_p$, respectively. Error
		bars in the calibration-noise scans represent one population standard
		deviation over 50 independent realizations.
}	
\label{fig9}
\end{figure}

 		The robustness analysis considers phase calibration noise, detuning
 		calibration noise, and finite pulse rise time. Independent Gaussian perturbations are added to the corresponding control parameter of each pulse. After the phase perturbation is applied, each phase is wrapped modulo $2\pi$. The perturbed detuning values are used directly, without clipping or projection back onto the discrete training grid. The phase-noise standard deviation is varied over
 		$\sigma_\phi/\pi\in\{0,0.01,0.02,0.05\}$, whereas the detuning-noise
 		standard deviation is varied over
 		$\sigma_\Delta\in\{0,0.01,0.02,0.05\}$. For each phase- and detuning-noise level, 50 independent realizations are evaluated. The plotted means and error bars in the calibration-noise scans are computed using the population mean and population standard	deviation over these realizations. Fixed NumPy random seeds are used for the phase-noise and detuning-noise scans to ensure reproducibility.
 		The finite-rise-time effect is modeled by interpolating between successive
 		pulse parameters. The phase is interpolated along the shortest angular
 		path, whereas the detuning is interpolated linearly. Each rise interval is
 		divided into eight numerical substeps, followed by the remaining constant
 		pulse-duration segment. The normalized rise-time fraction is varied over
 		$\tau_r/T_p\in\{0,0.05,0.10,0.20\}$. For Case A, the CFI under each robustness setting is evaluated at the boundary operating point $\eta_0=0$ using the right-sided-limit prescription described above. For Case B, the CFI is evaluated from the complete normalized observation model containing $p_0$, $p_1$, and $p_{\mathrm{loss}}$.

 For Case A, the three panels in Fig.~\ref{fig9}(a) show the dependence
 of the CFI on phase calibration noise, detuning calibration noise, and
 finite pulse rise time, respectively. The CFI decreases modestly with
 increasing phase-noise level and remains nearly unchanged over the tested
 detuning-noise range. The CFI changes only slightly at the smallest nonzero rise-time
 fraction and then decreases as the normalized rise-time fraction is
 increased further. These results characterize the sensitivity of the learned joint phase--detuning protocol for Case A to the tested control imperfections. For Case B, the three panels in Fig.~\ref{fig9}(b) show the corresponding robustness results for the learned joint phase--detuning control protocol in the effective non-Hermitian two-level model. The phase- and detuning-noise scans show only small changes in the CFI over the tested ranges, whereas the finite pulse rise time produces a more pronounced reduction. The results show how these control imperfections affect the sensing performance in the dissipative model. Together, the results in Fig.~\ref{fig9} characterize the response of the learned control protocols to calibration noise and finite modulation bandwidth in the two physical models.
    \begin{figure}
 	\centering
 	\hspace*{\fill}
 	\includegraphics[width=1.0\linewidth]{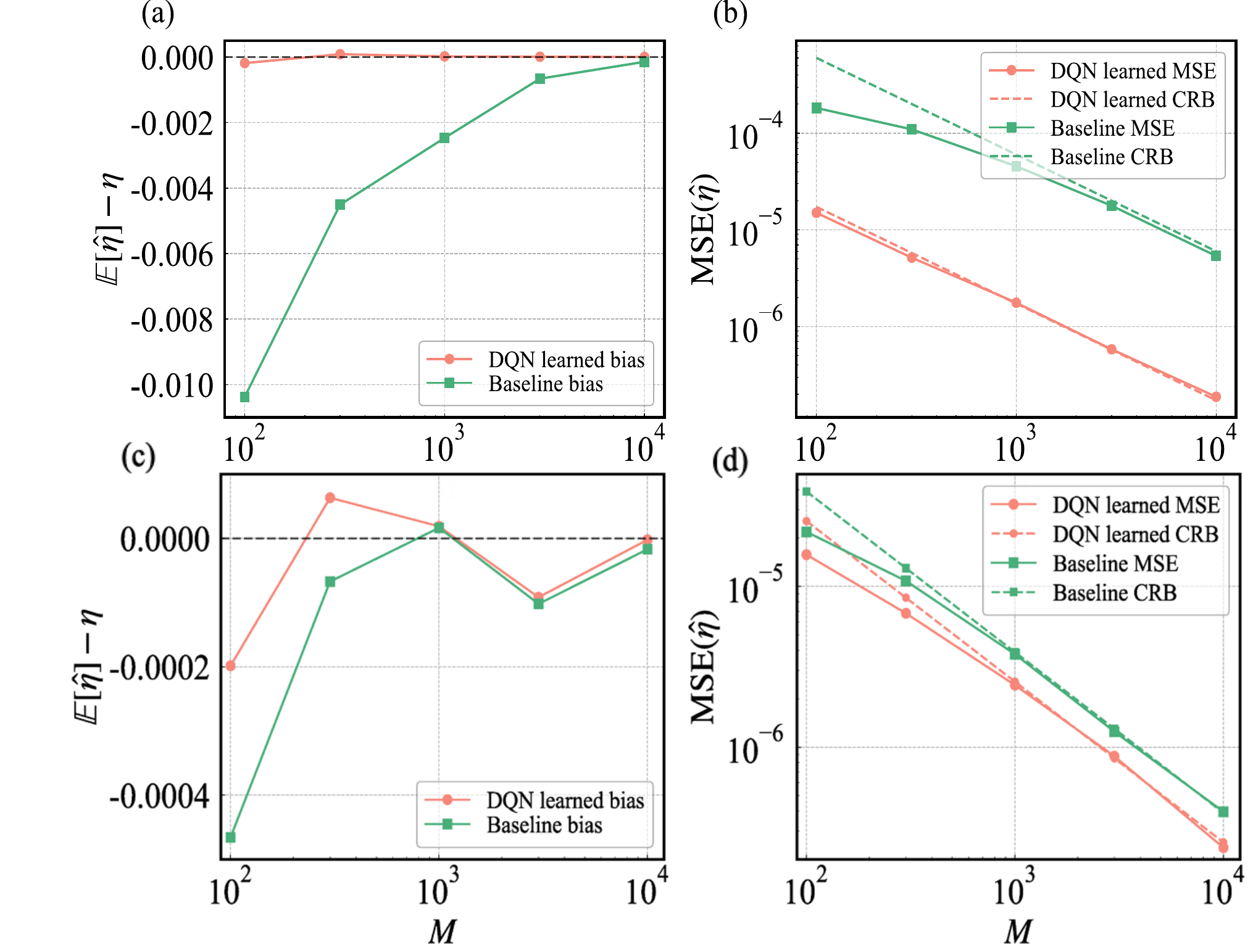}
 	\caption{
 		Finite-shot estimation performance of the DQN-optimized joint
 		phase--detuning protocol and the fixed-resonant baseline for Case A
 		[(a), (b)] and Case B [(c), (d)]. Panels (a) and (c) show the signed
 		estimation bias, while panels (b) and (d) show the empirical
 		mean-squared error (MSE) and the CFI-based Cram\'er--Rao reference curves. Solid and dashed curves denote the Monte Carlo
 		results and CFI-based references, respectively. Each point is obtained
 		from 500 independent Monte Carlo trials.
 	}
 	\label{fig10}
 \end{figure}

 \subsection{Finite-shot Measurement Performance}
 The CFI characterizes the local asymptotic sensitivity of a
 measurement protocol, whereas an experiment estimates the unknown
 parameter from a finite number of measurement outcomes. We therefore
 perform finite-shot maximum-likelihood-estimation (MLE) simulations
 for both the Hermitian three-level leakage model (Case A) and the
 effective non-Hermitian two-level model (Case B). In each case, the
 DQN-optimized joint phase--detuning pulse sequence is compared with
 the corresponding fixed-resonant baseline sequence, for which
 $\phi_k=0$ and $\Delta_k=0$ for every pulse. Within each model, the
 two protocols use the same initial state, Hamiltonian, pulse duration,
 ten-pulse sequence length, and observation model.
For Case A, projective measurements are performed in the basis
$\{|0\rangle,|1\rangle,|2\rangle\}$. For Case B, the complete
normalized observation model contains the outcomes
$\{0,1,\mathrm{loss}\}$, with
$p_{\mathrm{loss}}(\eta)=1-p_0(\eta)-p_1(\eta)$.
For each case $s\in\{A,B\}$, the finite-shot simulations are performed
at a fixed interior parameter point, denoted by
$\eta_{\mathrm{true}}^{(s)}$, within the likelihood-search interval
$[0.010,0.030]$. The same parameter point is used for the DQN-optimized
protocol and the fixed-resonant baseline at all measurement budgets.

 For compactness, we write
 	$\eta_s\equiv\eta_{\mathrm{true}}^{(s)}$. For Case A, we define
 	$\mathbf{n}_A=(n_0,n_1,n_2)$ and
 	$\mathbf{p}_A=(p_0,p_1,p_2)$.
 	For Case B, we define
 	$\mathbf{n}_B=(n_0,n_1,n_{\mathrm{loss}})$ and
 	$\mathbf{p}_B=(p_0,p_1,p_{\mathrm{loss}})$.
 	The measurement counts are sampled according to
 	\begin{equation}
 		\mathbf{n}_s
 		\sim
 		\operatorname{Mult}
 		\bigl(M;\mathbf{p}_s(\eta_s)\bigr),
 		\qquad s\in\{A,B\},
 	\end{equation} Here, $\operatorname{Mult}$ denotes the multinomial distribution.
 	For each case, protocol, and value of $M$, 500 independent Monte Carlo
 	data sets are generated using a fixed pseudorandom-number-generator
 	seed of 2026. For each case $s\in\{A,B\}$, the maximum-likelihood
 	estimate is obtained by maximizing
 	\begin{equation}
 		\log L_s(\eta)
 		=
 		\sum_{x\in\mathcal{X}_s}
 		n_x\log p_x^{(s)}(\eta),
 	\end{equation}
 	where
 	$\mathcal{X}_A=\{0,1,2\}$ and
 	$\mathcal{X}_B=\{0,1,\mathrm{loss}\}$.
 	The maximization is performed over a uniform grid of 4001 values in
 	$[0.010,0.030]$. The bias and MSE are calculated from the resulting
 	estimates, while the corresponding CFI-based Cram\'er--Rao reference
 	is evaluated as
 	\begin{equation}
 		\operatorname{CRB}_s(M)
 		=
 		\frac{1}
 		{M\mathcal{F}_C^{(s)}(\eta_{\mathrm{true}}^{(s)})}.
 	\end{equation}
 Because the finite-sample maximum-likelihood estimator may be biased,
 the CFI-based Cram\'er--Rao quantity is used as a local asymptotic
 precision reference rather than a strict lower bound on the empirical
 finite-sample MSE. Because the finite-shot parameter point used for
 each case lies in the interior of the likelihood-search interval, the
 corresponding CFI is evaluated with a central finite difference,
 \begin{equation}
 	\left.\partial_\eta p_x^{(s)}\right|_{\eta_s}
 	\approx
 	\frac{
 		p_x^{(s)}(\eta_s+\delta\eta)
 		-p_x^{(s)}(\eta_s-\delta\eta)}
 	{2\delta\eta},
 	\quad
 	\delta\eta=10^{-3}.
 \end{equation}
 	In evaluating the CFI, the probabilities
 	$p_x^{(s)}(\eta_{\mathrm{true}}^{(s)})$ are used in the denominators,
 	and outcomes satisfying
 	$p_x^{(s)}(\eta_{\mathrm{true}}^{(s)})\leq10^{-14}$
 	are omitted from the CFI sum to avoid numerical division by zero.
 	The plotted bias and MSE values are empirical averages over the 500
 	Monte Carlo trials, whereas the CFI-based reference curves are
 	determined directly from the corresponding CFI values. No error bars
 	are shown.

For Case A, the empirical signed estimation bias
	$\mathbb{E}[\hat{\eta}]-\eta_{\mathrm{true}}^{(A)}$ is shown in
	Fig.~\ref{fig10}(a). The DQN-optimized protocol maintains a small bias
	over the tested measurement budgets, whereas the fixed-resonant
	baseline exhibits a pronounced negative bias at small $M$ and
	approaches zero as $M$ increases. At $M=10000$, the biases are
	approximately $4.48\times10^{-6}$ for the DQN protocol and
	$-1.43\times10^{-4}$ for the fixed-resonant baseline. Fig.~\ref{fig10}(b)
	compares the corresponding empirical MSEs with the CFI-based reference
	curves. The DQN protocol gives a lower empirical MSE over the tested
	measurement budgets. At $M=10000$, the empirical MSEs are approximately
	$1.88\times10^{-7}$ for the DQN protocol and
	$5.38\times10^{-6}$ for the fixed-resonant baseline.
	For Case B, the empirical signed estimation bias
	$\mathbb{E}[\hat{\eta}]-\eta_{\mathrm{true}}^{(B)}$ is shown in
	Fig.~\ref{fig10}(c). The biases of both protocols remain close to zero
	over the tested measurement budgets. At $M=10000$, the biases are
	approximately $-2.23\times10^{-6}$ for the DQN protocol and
	$-1.70\times10^{-5}$ for the fixed-resonant baseline. As shown in
	Fig.~\ref{fig10}(d), the DQN protocol also gives a lower empirical MSE
	over the tested measurement budgets. At $M=10000$, the empirical MSEs
	are approximately $2.37\times10^{-7}$ for the DQN protocol and
	$3.96\times10^{-7}$ for the fixed-resonant baseline.
	At the largest measurement budget, the empirical MSEs are comparable
	in magnitude to the corresponding CFI-based reference values. Overall,
	the results characterize the finite-sample estimation performance of
	the two control protocols under the specified measurement and
	estimation procedures.

\begin{table*}
	\centering
	\caption{Numerical simulation parameters employed in the reinforcement-learning optimization}
	\label{tab:parameters}
	
	\renewcommand{\arraystretch}{1.2}
	
	\begin{tabular}{|>{\centering\arraybackslash}p{3.5cm}|>{\centering\arraybackslash}p{5.6cm}|>{\centering\arraybackslash}p{5.6cm}|}
		\hline
		\textbf{Parameter} & \textbf{Case A} & \textbf{Case B} \\
		\hline
		
		Physical model
		&
		Three-level $\Lambda$-type leakage model
		&
		Effective non-Hermitian two-level model
		\\
		\hline
		
		Unknown parameter
		&
		Leakage parameter $\eta$
		&
		Attenuation parameter $\eta$
		\\
		\hline
		
	Numerical propagator
		&
		$U=e^{-iH\Delta t}$
		&
		$U=e^{-iH_{\rm eff}\Delta t}$
		\\
		\hline

		Initial state
		& 
		$|0\rangle$
		&
		$|1\rangle$
		\\
		\hline
	
	Operating point 
		& $\eta\rightarrow0^{+}$
		& $\eta_0=0.025$
		\\
		\hline
		
	Training episodes
		& 600
		& 2500
	\\
	\hline
			
	Batch size
	& 64
	& 128
	 \\
	\hline
		Finite-difference step
		& \multicolumn{2}{c|}{$\delta\eta=0.001$} \\
		\hline
		
		Rabi frequency
		& \multicolumn{2}{c|}{$\Omega=\pi$} \\
		\hline
		
		Number of control pulses
		& \multicolumn{2}{c|}{$N=10$} \\
		\hline
		
		Pulse duration
		& \multicolumn{2}{c|}{$\Delta t=1$} \\
		\hline

		Optimizer
		& \multicolumn{2}{c|}{Adam} \\
		\hline

		Discount factor
		& \multicolumn{2}{c|}{0.98} \\
		\hline

		Exploration scheme
		& \multicolumn{2}{c|}{Exponential $\varepsilon$-greedy ($1.0\rightarrow0.05$)} \\
		\hline
		
		Random seed
		& \multicolumn{2}{c|}{2016-2025} \\
		\hline
		
	Phase action set
		& \multicolumn{2}{c|}{ $[0,2\pi)$} \\
		\hline
		
		Detuning action set
		&
		Joint control: $\{-1,0,1\}$
		&
	Joint control: $\{-3,-2,-1,0,1,2,3\}$
		\\
		\hline
		
		Learning rate
		&
		0.005
		&
		0.003
		\\
		\hline
		
	\end{tabular}
	
	\vspace{2mm}
	\footnotesize
	\raggedright  
	\label{Table II}	
\end{table*} 
\begin{table*}
\centering
\caption{Optimized pulse sequences discovered by the trained DQN}
\label{tab:pulse_sequence}

\renewcommand{\arraystretch}{1.3}

\begin{tabularx}{\textwidth}{|c|c|>{\raggedright\arraybackslash}X|}
	\hline
	\textbf{Model} & \textbf{Strategy} & \multicolumn{1}{c|}{\textbf{Representative learned pulse sequence $\{(\phi_k,\Delta_k)\}_{k=1}^{10}$ }} \\
	\hline
	
	\multirow{2}{*}{Case A}
	& Phase-only
	& $\{\tfrac{1}{7}\pi,\,\tfrac{6}{7}\pi,\,\tfrac{13}{7}\pi,\,\tfrac{5}{7}\pi,\,\tfrac{12}{7}\pi,\,
	\tfrac{10}{7}\pi,\,\tfrac{3}{7}\pi,\,\tfrac{5}{7}\pi,\,\tfrac{12}{7}\pi,\,\tfrac{13}{7}\pi\}$ \\
	\cline{2-3}
	& Phase-detuning
	& $\{(\pi/2,-1),\,(0,0),\,(\pi,0),\,(0,0),\,(\pi,0),\,
	(0,0),\,(\pi,0),\,(0,0),\,(\pi,0),\,(3\pi/2,0)\}$ \\
	\hline
	
	\multirow{2}{*}{Case B}
	& Phase-only
	&  $\{0,\,\tfrac{28}{15}\pi,\,\tfrac{2}{3}\pi,\,\tfrac{28}{15}\pi,\,\tfrac{16}{15}\pi,\,0,\,\tfrac{14}{15}\pi,\,0,\,\tfrac{16}{15}\pi,\,0\}$  \\
	\cline{2-3}
	& Phase-detuning
	& $\{( \tfrac{11}{8}\pi,\,2),\,(\pi,\,3),\,(\tfrac{7}{8}\pi,\,3),\,(\tfrac{7}{8}\pi,\,3),\,(\pi,\,3),\,(\pi,\,3),\,(0,\,-3),\,(\pi,\,3),\,(\pi,\,3),\,(\pi,\,3)\}$ \\
	\hline
\end{tabularx}

\vspace{2mm}
\footnotesize
\raggedright  
\label{Table III}	
\end{table*}  
\section{CONCLUSION}
\label{sec5}
In summary, we present a reinforcement learning-based quantum sensing algorithm for high-precision estimation of level leakage. As a major
error source in quantum state control, level leakage causes population
transfer out of the computational subspace and can give rise to
non-unitary and non-trace-preserving dynamics in effective reduced
descriptions. Our method uses pulse-level reinforcement learning to optimize control protocols and improve the measurement-model-specific CFI for leakage estimation when the system Hamiltonian and the leakage-signal generator do not commute.
We adopt a pulse-level parameter modulation strategy for investigations
on Hermitian three-level and effective non-Hermitian two-level systems,
thereby avoiding explicit gate decomposition associated with
conventional gate-based control. For the pulse sequences discovered by the DQN, the CFI of the joint
	phase--detuning protocols closely approaches the corresponding QFI in
	both models. In Case A, the optimized CFI exhibits an approximately
	quadratic dependence on the pulse number $N$ over the considered range.
	In Case B, at $N=10$, joint phase--detuning control yields a CFI
	approximately $23\%$ higher than that obtained with phase-only control,
	while the optimized CFI increases approximately linearly with $N$. Moreover, further precision improvements may be possible by modulating more pulse parameters. These results show that reinforcement learning can effectively optimize
	pulse-level controls and improve leakage-sensing performance under the
	specified physical and measurement conditions. The results demonstrate the applicability of the proposed framework to
		the two model systems considered here. The framework may provide a useful
		approach for leakage characterization in near-term quantum sensing
		experiments and motivate further studies of leakage-aware control for
		fault-tolerant quantum information processing.

\section*{acknowledgements}
This work was supported by the National Natural Science Foundation of China (Grant Nos. 62471001, 12475009, 12075001, and 12175001), the Natural Science Research Project in Universities of Anhui Province (Grant No. 2024AH050068), Anhui Provincial Key Research and Development Plan (Grant No. 2022b13020004), and Anhui Provincial Natural Science Foundation (Grant No. 1508085QF139).

\section*{data availability}

There are no publicly available research data or software supporting this paper. Requests for further information or data should be sent to the authors. 
\section*{Appendix A: Simulation and Training Parameters}
	The simulation parameters and reinforcement-learning hyperparameters used throughout this work are summarized in Table.~\ref{Table II}. The table provides the key settings for both the Hermitian three-level leakage model (Case A) and the non-Hermitian two-level model (Case B), including the initial states, nominal and finite-difference parameters, numerical propagators, Rabi frequency, control-pulse number and duration, DQN optimization settings, exploration scheme, random seed, action spaces for phase and detuning control, and learning rates.

\section*{Appendix B: Optimized pulse sequences discovered by the trained DQN}
	The representative optimized pulse sequences obtained from the trained DQN are summarized in Table.~\ref{Table III} for the two physical models and the different control strategies considered in this work. For phase-only control, each entry specifies the pulse phase $\phi$, whereas for joint phase--detuning control, each pulse is represented by a pair $(\phi,\Delta)$. Each sequence contains $N=10$ control pulses and corresponds to a representative solution obtained after training.
\end{document}